\documentclass{sig-alternate-2013}
\newfont{\mycrnotice}{ptmr8t at 7pt}
\newfont{\myconfname}{ptmri8t at 7pt}
\let\crnotice\mycrnotice%
\let\confname\myconfname%

\usepackage{hyperref}
\usepackage{breakurl}
\usepackage{graphicx}
\usepackage{xcolor}

\usepackage{lgrind}
\usepackage{subcaption}
\usepackage{caption}
\usepackage[linesnumbered,lined,ruled,noend,algosection]{algorithm2e}
\usepackage{chngcntr}
\counterwithin{figure}{section}
\counterwithin{table}{section}

\def\ojoin{\setbox0=\hbox{$\Join$}%
  \rule[0.1ex]{.27em}{.4pt}\llap{\rule[1.3ex]{.27em}{.4pt}}}
\def\leftouterjoin{\mathbin{\ojoin\mkern-5.8mu\Join}}
\newtheorem{definition}{Definition}
\newtheorem{prop}{Property}
\newtheorem{thm}{Lemma}
\counterwithin{thm}{section}
\counterwithin{definition}{section}
\counterwithin{prop}{section}

\begin{document}

\title{{\ttlit Left Bit Right}: For SPARQL Join Queries with OPTIONAL Patterns (Left-outer-joins)}
\numberofauthors{1} 
\author{
\alignauthor
Medha Atre\titlenote{The initial part of this work was completed when
the author was affiliated to Rensselaer Polytechnic Institute and University of Pennsylvania.}\\
        \affaddr{Pune, India}\\
       \email{medha.atre@gmail.com}
}

\maketitle
\begin{abstract}
SPARQL basic graph pattern (BGP) (a.k.a. SQL inner-join) query optimization
is a well researched area. However, optimization of OPTIONAL pattern
queries (a.k.a. SQL left-outer-joins)
poses additional challenges, due to the restrictions on the \textit{reordering}
of left-outer-joins. The occurrence of such queries
tends to be as high as 50\% of the total queries (e.g., DBPedia query logs). 

In this paper, we present \textit{Left Bit Right} (LBR), a technique for
\textit{well-designed} nested BGP and OPTIONAL pattern queries.
Through LBR, we propose a novel method to represent such queries 
using a graph of \textit{supernodes}, which is used to aggressively
prune the RDF triples, with the help of compressed indexes.
We also propose novel optimization strategies --
first of a kind, to the best of our knowledge
-- that combine together the characteristics of \textit{acyclicity}
of queries, \textit{minimality}, and \textit{nullification},
\textit{best-match} operators.
In this paper, we focus on OPTIONAL patterns without UNIONs or FILTERs,
but we also show how UNIONs and FILTERs can be handled with our technique
using a \textit{query rewrite}.
Our evaluation on RDF graphs of up to and over one billion triples,
on a commodity laptop with 8 GB memory, shows that LBR can process \textit{well-designed}
low-selectivity complex queries up to 11 times faster compared to
the state-of-the-art RDF column-stores as Virtuoso and MonetDB, and for highly selective
queries, LBR is at par with them.
\end{abstract}


\section{Introduction} \label{sec:intro}
Resource Description Framework (RDF) \cite{rdf}
is a widely accepted standard for representing \textit{semantically linked} data on the web, and
SPARQL \cite{sparql}  is a standard query language for it.
RDF data is a directed edge-labeled multi-graph, where each unique edge (S P O) is called 
a \textit{triple} -- P is the label on the edge from the node S to node O.

SPARQL query language, like SQL, provides various syntactic constructs to
form \textit{structured} queries for RDF graphs. Two types of queries of our interest
are \textit{Basic Graph Pattern} (BGP) and \textit{OPTIONAL} pattern (OPT) queries.
Any SPARQL BGP query can be methodically translated into an equivalent SQL inner-join
query \cite{sparqlalg}, and an OPT pattern query can be methodically
translated into an SQL left-outer-join query \cite{opttosql}.
Like SQL, the SPARQL grammar allows nested OPT queries, i.e.,
a query composed of an intermix of BGPs and OPT patterns.
Since SPARQL BGP queries are same as SQL inner-joins (denoted by symbol $\Join$)
and OPT queries are same as SQL left-outer-joins (denoted by $\leftouterjoin$),
we will use these terms, acronyms, and symbols interchangeably in the rest of the text.
BGP and OPT queries tend to be performance intensive, especially for very large RDF data.
An extensive amount of research has gone in the optimization of SQL inner-join queries.
The database and semantic web communities have taken these optimizations further along with the novel ideas
of indexing and BGP query processing over RDF data \cite{bitmatwww10,triad,hrdf3x,rdf3x,swans,hexastore,triplebit}.

\textbf{Why OPT queries?}
RDF is a semi-structured data, and adherence of RDF ``instance data'' to its Ontology specification (a.k.a. schema),
and in turn its completeness, is not always enforced especially for the data published on the web (e.g., DBPedia \cite{dbpedia}),
and that which is compiled from many diverse sources of RDF graphs (e.g., Linked Open
Data\footnote{\scriptsize{\url{http://linkeddata.org/}}}).
This makes OPT queries a crucial tool for the end users.
For example, consider an RDF network describing movie and TV sitcom/soap actors.
Not all the actors may have their contact info, such as email and telephone numbers listed. Then an OPT query
(\textbf{Q1}) as given below fetches \textbf{all} the \textit{actors} with their respective
\textit{name} and \textit{address}. Along with that, it gets \textit{email} and
\textit{tele} numbers of those who have them listed, and
for the rest it marks \textit{email} and \textit{tele} values by \textit{NULL}s.

\vspace{1mm}
\begin{lgrind}
{\small
\Head{}
\L{\LB{\K{SELECT}_?\V{actor}_?\V{name}_?\V{addr}_?\V{email}_?\V{tele}_\K{WHERE}_\{}}
\L{\LB{}\Tab{5}{?\V{actor}_:\V{name}_?\V{name}_.}}
\L{\LB{}\Tab{5}{?\V{actor}_:\V{address}_?\V{addr}_.}}
\L{\LB{}\Tab{5}{\K{OPTIONAL}_\{}}
\L{\LB{}\Tab{7}{?\V{actor}_:\V{email}_?\V{email}_.}}
\L{\LB{}\Tab{7}{?\V{actor}_:\V{telephone}_?\V{tele}_.\}\}}}
}
\end{lgrind}

Query logs of SPARQL endpoints on the web indeed concur with this intuition, e.g.,
DBPedia query logs show as high as 50\% occurrence of OPT queries \cite{usewod11,swim},
with as many as eight OPT patterns in a query.
These statistics make OPT queries a non-negligible component of SPARQL ``join'' query optimization.
Now consider a modified version (\textbf{Q2}) of the previous query:

\vspace{1mm}
\begin{lgrind}
{\small
\Head{}
\L{\LB{\K{SELECT}_?\V{friend}_?\V{sitcom}_\K{WHERE}_\{}}
\L{\LB{}\Tab{2}{:\V{Jerry}_:\V{hasFriend}_?\V{friend}_.}}
\L{\LB{}\Tab{2}{\K{OPTIONAL}_\{}}
\L{\LB{}\Tab{4}{?\V{friend}_:\V{actedIn}_?\V{sitcom}_.}}
\L{\LB{}\Tab{4}{?\V{sitcom}_:\V{location}_:\V{NewYorkCity}_.\}\}}}
}
\end{lgrind}
\vspace{1mm}

This query asks for all friends of \textit{:Jerry} that have acted in a sitcom located in the \textit{:NewYorkCity}.
In this query, let (\textit{:Jerry :hasFriend ?friend}) be $tp_1$, (\textit{?friend :actedIn ?sitcom})
$tp_2$, and (\textit{?sitcom :location :NewYorkCity}) $tp_3$.
Then the query can be expressed as $(Query = tp_1 \leftouterjoin (tp_2 \Join tp_3))$, where
$tp_1$ forms a BGP (say $P_1$) with only one triple pattern, and $(tp_3 \Join tp_4)$ forms another BGP (say $P_2$).

BGP queries ($\Join$) are \textit{associative} and \textit{commutative},
i.e., a change in the order of triple patterns and joins between them
does not change the final results.
But for a nested OPT query, $\leftouterjoin$ operator is not associative and commutative,
e.g., in the case of \textbf{Q2} above, left-outer-join between
$tp_1$ and $tp_2$ cannot be performed before the inner-join $P_2 = (tp_2 \Join tp_3)$.
This limits the number of query plans an optimizer can consider.

Let us assume that this actor network has thousands of actors, and many of them have acted
in sitcoms located in the \textit{:NewYorkCity}.
But \textit{:Jerry} has just two friends,
\textit{:Larry} and \textit{:Julia}, and among them only \textit{:Julia}
has acted in \textit{:Seinfeld}, which has \textit{:NewYorkCity} as the location.
So the final results of this query are just two with the respective values
of \textit{?friend} and \textit{?sitcom} set as 
\{(\textit{:Larry, NULL}), (\textit{:Julia, :Seinfeld})\}.
However, the inner-join $P_2 = tp_2 \Join tp_3$ has to be evaluated before the left-outer-join
$P_1 \leftouterjoin P_2$, due to the restrictions on the
reorderability of left-outer-joins.
Since there are several actors who have acted in the sitcoms located in \textit{:NewYorkCity},
these two triple patterns have a \textit{low selectivity}\footnote{\scriptsize{Selectivity of a triple
pattern is high if it has fewer number of triples associated with it and vice versa.}}, which increases
the evaluation time and cost of their inner-join.
To overcome such a limitation, for conventional databases, Rao et al \cite{rao2,rao1}, and
Galindo-Legaria, Rosenthal \cite{galindo-legaria2}
have proposed ways of reordering nested inner and left-outer joins using additional operators as
\textit{nullification} and \textit{best-match} \cite{rao1} (\textit{Generalized Outerjoin} in
\cite{galindo-legaria2}).

On this background, we propose \textit{Left Bit Right} (LBR), and make
the following main contributions.

$\bullet$ We propose a novel way to represent a nested BGP-OPT query
 with a graph of \textit{supernodes} (Section \ref{sec:qgrconstr}).
 
$\bullet$
We extend the previously known properties of 
 \textit{acyclicity} of queries, \textit{minimality} of triples, and \textit{nullification}, \textit{best-match} operations,
 to propose novel optimization strategies
  -- first of a kind, to the best of our knowledge -- for acyclic and cyclic \textit{well-designed} OPT queries
  (Sections \ref{sec:semijloj}, \ref{sec:cyclic}, \ref{sec:qproc}).
  We mainly focus on the ``join'' component of SPARQL, i.e., OPT patterns without
  UNIONs, FILTERs, or Cartesian products. Nevertheless, in Section \ref{sec:discuss} we show
  how these constructs can be handled using our technique.

$\bullet$ Finally, we show LBR's performance using a commodity laptop of 8 GB memory
 over three popular RDF datasets, UniProt \cite{uniprot}, LUBM \cite{lubm}, and
 DBPedia \cite{dbpedia}, with up to and over a billion triples,
 in comparison with the state-of-the-art RDF column-stores like Virtuoso v7.1.0 and MonetDB v11.17.21.
 Through our evaluation on queries with varying degrees of selectivity, complexity, and running times,
 we show that for complex, low selectivity ``\textit{well-designed}'' queries,
 LBR is up to 11 times faster than Virtuoso and MonetDB,
 and for highly selective queries, it is at par with them (Section \ref{sec:eval}).

\section{Query Graph of Supernodes} \label{sec:qgrconstr}
An OPT query with an intermix of Basic Graph Patterns (BGPs $\Join$)
and OPT patterns ($\leftouterjoin$) establishes restrictions
on the order of join processing, e.g., \textbf{Q1}, \textbf{Q2} in Section \ref{sec:intro}.
In this section we introduce a novel idea of constructing a \textit{graph of supernodes} (GoSN) to
capture the nesting of OPT patterns in a query. 
GoSN makes an important part of our query processing
techniques discussed in Sections \ref{sec:optstrategy} and \ref{sec:qproc}.

\subsection{GoSN Construction} \label{sec:qgrconstr1}
\textbf{Supernodes:}
In a SPARQL OPT pattern of the form $(P_1 \leftouterjoin P_2)$, $P_1$
may in turn have nested BGPs and OPT patterns inside it, e.g., $P_1 = (P_3 \leftouterjoin P_4)$.
$P_2$ may have nested BGPs and OPT patterns inside it too, or either of $P_1$ and $P_2$ can be OPT-free.
Generalizing it, if a pattern $P_i$ does not have any OPT pattern nested inside it,
we call $P_i$ to be an \textit{OPT-free Basic Graph Pattern}.
From a given nested OPT query, first we extract all such OPT-free BGPs, and construct 
a \textit{supernode}  ($SN_i$) for each $P_i$.
The triple patterns (TPs) in $P_i$ are \textit{encapsulated} in $SN_i$.

\begin{figure}[!h]
        \centering
        \begin{subfigure}[b]{0.32\textwidth}
                \includegraphics[width=0.9\textwidth, height=0.6in]{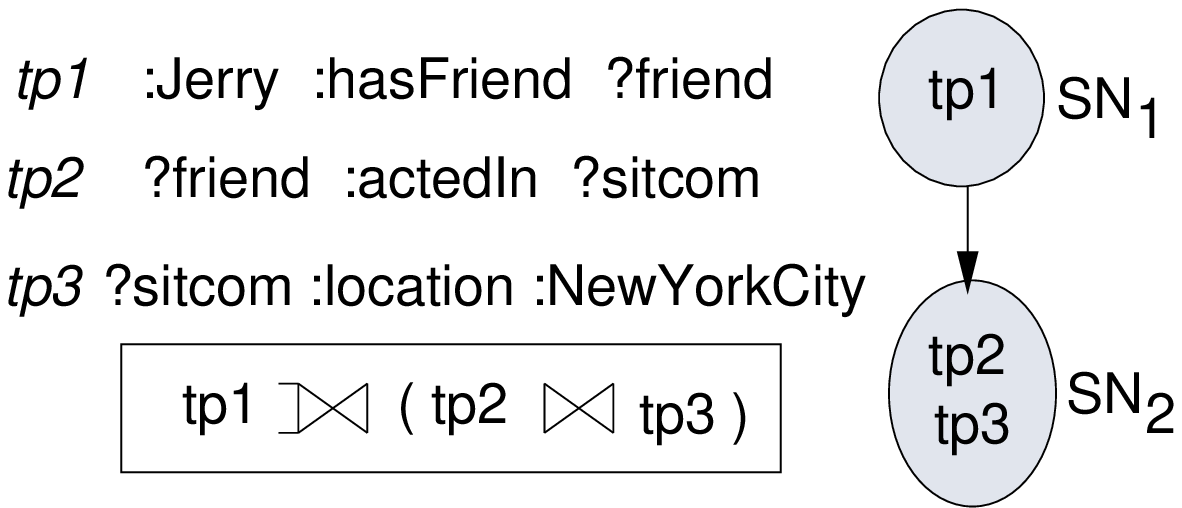}
                \caption{}\label{fig:qgraph}
        \end{subfigure}%
        \begin{subfigure}[b]{0.17\textwidth}
                \includegraphics[width=0.9\textwidth, height=0.6in]{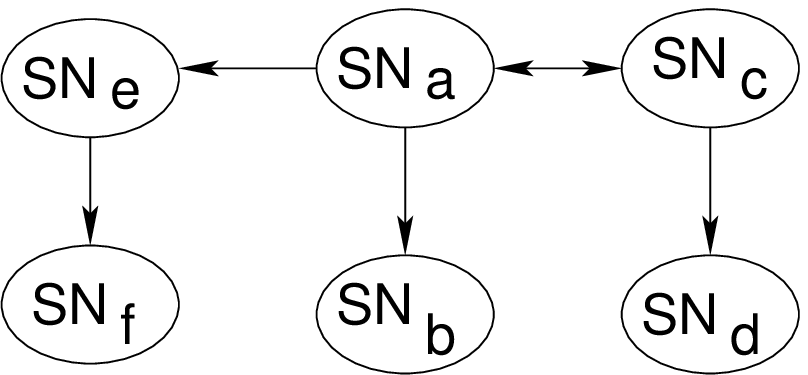}
                \caption{} \label{fig:qgrex}
        \end{subfigure}
        \captionsetup{singlelinecheck=off}
        \caption[.]{GoSN for -- (a) Q2 in Section \ref{sec:intro}, (b)
        $\mathbf{((P_a \leftouterjoin P_b) \Join (P_c \leftouterjoin P_d)) \leftouterjoin (P_e \leftouterjoin P_f)}$}
\end{figure}

Next, we serialize a nested OPT query using its OPT-free BGPs, 
$\Join$ (inner-join), $\leftouterjoin$ (left-outer-join) operators, and proper parentheses.
E.g., we serialize \textbf{Q2} in Section \ref{sec:intro} as ($P_1 \leftouterjoin P_2$),
where $P_1$ and $P_2$ are OPT-free BGPs, $SN_1$ of $P_1$ encapsulates just $tp_1$,
and $SN_2$ of $P_2$ encapsulates $tp_2$ and $tp_3$ (see Figure \ref{fig:qgraph}). 

\textbf{Unidirectional edges:}
From the serialized query, we consider each OPT pattern of type $P_m \leftouterjoin P_n$.
$P_m$ or $P_n$ may have nested OPT-free BGPs inside them.
Using the serialized-parenthesized form of the query,
we identify the \textit{leftmost} OPT-free BGPs nested inside $P_m$
and $P_n$ each. E.g., if $P_m = ((P_a \leftouterjoin P_b) \Join (P_c \leftouterjoin P_d))$,
and $P_n = (P_e \leftouterjoin P_f)$,
$P_a$ and $P_e$ are the \textit{leftmost} OPT-free BGPs in $P_m$ and $P_n$ respectively,
and $SN_a$ and $SN_e$ are their respective supernodes. We add a directed edge $SN_a \rightarrow SN_e$.
If either $P_m$ or $P_n$ does not nest any OPT-free BGPs inside it,
we treat the very pattern as the \textit{leftmost} for adding a directed edge.
With this procedure, we can treat OPT patterns in a query in any order.
But for all practical purposes, we start from the \textit{innermost} OPT patterns,
and recursively go on considering the outer OPT patterns using the
parentheses in the serialized query.
E.g., if a serialized query is ($(P_a \leftouterjoin P_b)$ $\Join$
$(P_c \leftouterjoin P_d)$) $\leftouterjoin$ ($P_e \leftouterjoin P_f$), with $P_{a}...P_{f}$ as OPT-free BGPs,
we add directed edges as follows: (1) $SN_a \rightarrow SN_b$, (2) $SN_c \rightarrow SN_d$,
(3) $SN_e \rightarrow SN_f$, (4) $SN_a \rightarrow SN_e$.

\textbf{Bidirectional edges:}
Next we consider each inner-join of type $P_x \Join P_y$ in a serialized query.
If $P_x$ or $P_y$ has nested OPT-free BGPs inside,
we add a bidirectional edge between the supernodes of \textit{leftmost} OPT-free BGPs.
E.g., if $P_x = (P_a \leftouterjoin P_b)$, and $P_y =(P_c \leftouterjoin P_d)$,
we add a bidirectional edge $SN_a \leftrightarrow SN_c$.
If $P_x$ or $P_y$ does not nest any OPT-free BGPs inside it,
we consider the very pattern to be the \textit{leftmost}
for adding a bidirectional edge.
We add bidirectional edges starting from the \textit{innermost} inner-joins ($\Join$)
using the parentheses in the serialized query,
and recursively go on considering the outer ones, until no more bidirectional edges can be added.
Considering the same example given under unidirectional edges,
we add a bidirectional edge between $SN_a \leftrightarrow SN_c$.
The \textit{graph of supernodes} (GoSN) for this example is shown in
Figure \ref{fig:qgrex}. 

Thus we completely capture the nesting of BGPs and OPT patterns in a query using this GoSN, and 
establish an order among the supernodes, which is described in Section \ref{sec:nomen}.

\subsection{Nomenclature}\label{sec:nomen}
In this section, we highlight the nomenclature that we use in the context
of GoSN, OPT patterns, and RDF graphs.

\textbf{Master-Slave:} In an OPT pattern $P_c \leftouterjoin P_d$,
we call pattern $P_c$ to be a \textit{master} of $P_d$, and $P_d$ a \textit{slave} of $P_c$.
This master-slave relationship is \textit{transitive},
i.e., if a supernode $SN_f$ is reachable from another supernode $SN_c$ by following \textit{at least}
one unidirectional edge in GoSN, then $SN_c$ is called a master of $SN_f$
(see Figure \ref{fig:qgrex}).

\textbf{Peers:} We call two supernodes to be peers if they are 
connected to each other through a bidirectional edge, or they can be reached
from each other by following \textit{only} bidirectional edges in GoSN, e.g.,
$SN_a$ and $SN_c$ in Figure \ref{fig:qgrex}.

\textbf{Absolute masters:} Supernodes that are \textit{not} reachable from any other
supernode through a path involving \textit{at least one} unidirectional edge are called the
\textit{absolute masters}, e.g., $SN_a$ and $SN_c$ in Figure \ref{fig:qgrex}
are absolute masters.

These master-slave, peer, and absolute master nomenclatures and
relationships apply to any triple patterns enclosed within the
respective supernodes too.

\textbf{Well-designed patterns:}
As per the definition given by P\'{e}rez et al \cite{perez2},
a \textit{well-designed} OPT query is --
for every subpattern of type $P' = P_k \leftouterjoin P_l$ in the query, if
a join variable ``?j'' in $P_l$ appears outside $P'$,
then ``?j'' also appears in $P_k$. A query that violates
this condition is said to be \textit{non-well-designed}.
In this paper, we have focused on well-designed queries,
because they occur most commonly for RDF graphs,
and remain unaffected by the difference between SPARQL and SQL algebra
over treatment of NULLs.
Nevertheless, we discuss \textit{non-well-designed} queries in
Appendices \ref{apdx:wdpat} and \ref{apdx:nulltreat}
for the completeness of the text.

\textbf{NULLs and blank nodes:}
Unlike relational tables, RDF graphs do not have NULLs. 
Note that NULLs represent \textit{non-existence} of entities, whereas ``blank nodes'' in
an RDF graph have ``blank node identifiers'' to represent entities without distinct
URIs\footnote{\scriptsize{\url{www.w3.org/TR/REC-rdf-syntax/\#section-Syntax-blank-nodes}}},
and in SPARQL queries, they are treated similar to the entities that have URIs.
An OPT query may generate NULLs, and 
joins over NULLs happen only in \textit{non-well-designed} patterns.
SPARQL and SQL algebra handles joins over NULLs differently, and 
Appendix \ref{apdx:nulltreat} elaborates on this issue.
Well-designed patterns, however, remain unaffected by this -- which are the focus of this paper. 

\section{Optimization Strategies} \label{sec:optstrategy}
\begin{figure}[h]
    \centering
       \includegraphics[width=3.4in, height=1.55in]{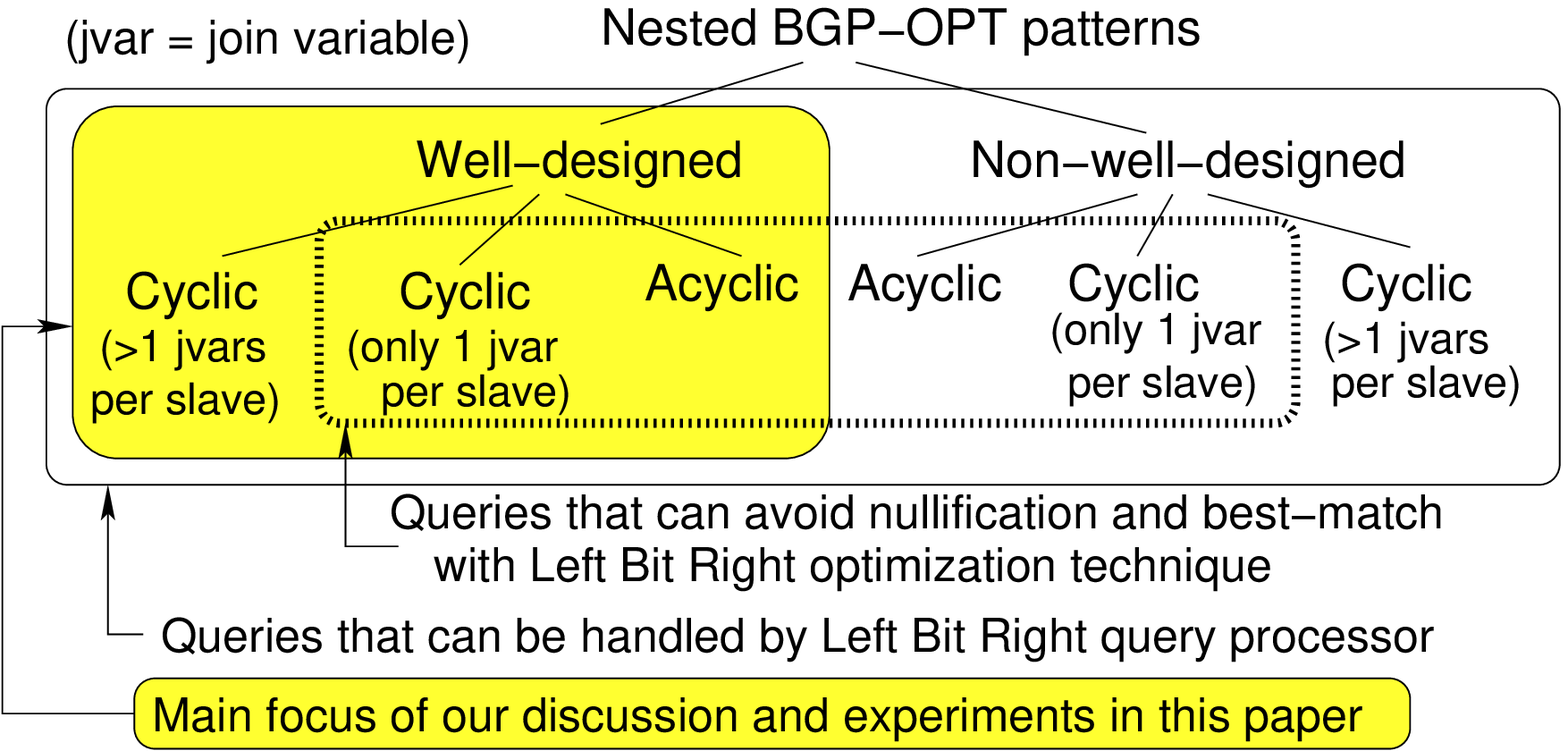}
       \caption{Classification of OPT queries} \label{fig:optqclass}
\end{figure}

In Figure \ref{fig:optqclass} we point out 
that LBR can process all the nested BGP-OPT queries, but not
all of them can avoid \textit{nullification} and \textit{best-match},
and we mainly focus on the \textit{well-designed} queries in this paper.
Our technique can be applied to \textit{non-well-designed} queries too,
but we have not focused on them due to lack of evidence of such queries in practice,
disparity between pure SPARQL and SPARQL-over-SQL engines for joins over NULLs
(refer to Appendices \ref{apdx:wdpat} and \ref{apdx:nulltreat}),
and space limitations. 

In Section \ref{sec:bestm}, briefly ignoring GoSN,
we describe the conceptual foundation of our technique by
combining the properties of \textit{nullification}, \textit{best-match}, \textit{acyclicity} of queries,
and \textit{minimality} of triples.
Then in Section \ref{sec:semijloj} we present our optimization strategies
to ``prune'' the triples for \textit{acyclic} well-designed patterns using GoSN,
and in Section \ref{sec:cyclic} we discuss pruning for \textit{cyclic} well-designed patterns.
For the discussion in Sections \ref{sec:bestm}, \ref{sec:semijloj}, \ref{sec:cyclic}
we do not take into consideration the underlying indexes on RDF
graphs, because the optimization strategies are agnostic to them.
Then in Section \ref{sec:bitmat} we describe our indexes, and in Section \ref{sec:qproc}
we show how to prune the triples using our indexes, and generate the final query results
using \textit{multi-way-pipelined} join for both, acyclic as well as cyclic well-designed queries.

\subsection{Preliminaries} \label{sec:bestm}
Typically, for a pairwise join processing plan, if nested inner and left-outer joins
are reordered, \textit{nullification} and \textit{best-match} (a.k.a.
\textit{minimum-union}) operations are required. We explain them here with a brief
example for the completeness of the text, and refer the reader to
\cite{galindosigmod, rao1} for the details.

\begin{figure}[!h]
    \centering
        \includegraphics[width=3.3in,height=2in]{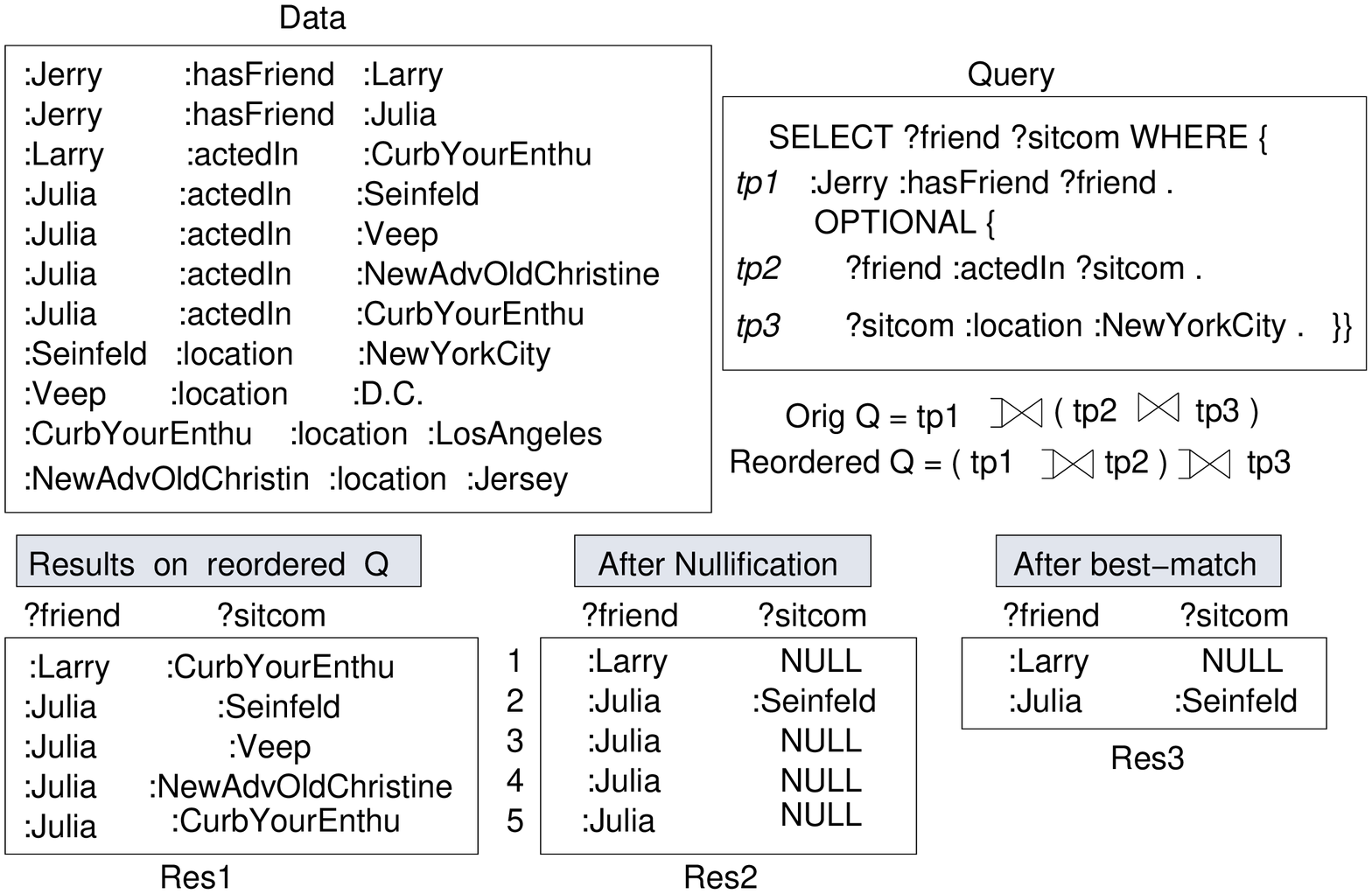}
       \caption{Nullification and best-match example} \label{fig:nullbest}
\end{figure}

\textbf{Nullification, Best-match:}
Consider the same query given in Figure \ref{fig:qgraph}, along with the sample
data associated with it in Figure \ref{fig:nullbest}.
\textit{:NewYorkCity} has been the location for a lot of American sitcoms,
and a lot of actors have acted in them (they are not shown in the sample data for conciseness).
But, among all such actors, \textit{:Jerry} has only two friends, \textit{:Julia} and \textit{:Larry}.
Hence, $tp_1$ is more \textit{selective} than $tp_2$ and $tp_3$. A left-outer-join
reordering algorithm as proposed in \cite{galindo-legaria2,rao1}
will typically reorder these joins as $(tp_1 \leftouterjoin tp_2) \leftouterjoin tp_3$.
Due to this reordering, all four sitcoms that \textit{:Julia} has acted in show up as bindings of
\textit{?sitcom} (see Res1 in Fig. \ref{fig:nullbest}),
although only \textit{:Seinfeld} was located in the \textit{:NewYorkCity}.
To fix this, \textit{nullification} operator is used, which
ensures that variable bindings across the reordered joins are consistent with
the original join order in the query (see Res2 in Figure \ref{fig:nullbest}).

We can see that the \textit{nullification} operation caused results that are \textit{subsumed}
within other results. A result $r_1$ is said to be subsumed within another result $r_2$ ($r_1 \sqsubset r_2$), if 
for every non-null variable binding in $r_1$, $r_2$ has the same binding, and $r_2$ has more
non-null variable bindings than $r_1$. Thus results 3--5 in Res2 are subsumed within result 2.
The \textit{best-match} operator removes all the subsumed results (see Res3).
Final results of the query are given as
\textit{best-match(nullification($(tp_1 \leftouterjoin tp_2) \leftouterjoin tp_3$))}.

\textbf{Semi-join} ($\ltimes$) is a well-known concept in databases, with which triples
associated with a triple pattern (TP) can be removed due to restrictions
on the variable bindings coming from another TP, without actually performing a join
between the two.
$tp_2 \ltimes_{?j} tp_1 = \{t\ |\ t \in tp_2, t.?j \in (\pi_{?j}(tp_1) \cap \pi_{?j}(tp_2))\}$.
Here $t$ is a triple matching $tp_2$, and $t.?j$ is a variable binding (value)
of variable $?j$ in $t$.
After this semi-join, $tp_2$ is left with only triples
whose $?j$ bindings are also in $tp_1$, and all other triples are removed. 
Now let triple patterns, $tp_1, tp_2, ...tp_n$ all share join variable \textit{?j}.
Then we define a \textbf{clustered-semi-join} over them as follows.
\begin{definition}
A clustered-semi-join(?j, \{$tp_1, tp_2,$ ...., $tp_n$\})
is performed as follows: Let $\mathcal{J} = \{\pi_{?j}(tp_1) \cap ... \cap \pi_{?j}(tp_n)\}$.
For each triple pattern $tp_i, 1 \leq i \leq n$, $tp_i = \{t\ |\ t \in tp_i, t.?j \in \mathcal{J}\}$.
\end{definition}

This definition shows that a clustered-semi-join is nothing but many 
semi-joins performed together for TPs that share a join variable.

\textbf{Example-1:} Now let us evaluate the query in Figure \ref{fig:nullbest} using
semi-joins and clustered-semi-joins.
We do a semi-join $tp_2 \ltimes_{?friend} tp_1$, because $tp_1$ does
a left-outer-join with $tp_2$ over \textit{?friend}.
That keeps only \textit{:Larry} and \textit{:Julia}
bindings of \textit{?friend} in $tp_2$, and removes any other bindings, and in turn triples
generating those bindings from $tp_2$ (they are not shown in the figure for conciseness).
Followed by it, we do a \textit{clustered-semi-join(?sitcom, \{$tp_2, tp_3$\})}.
That removes \textit{:CurbYourEnthu}, \textit{:Veep}, and \textit{:NewAdvOldChristine} bindings of \textit{?sitcom},
and the respective triples from $tp_2$. Notice that this clustered-semi-join
also removes the \textit{:Larry} binding of \textit{?friend} from $tp_2$ as a \textit{ripple effect}
of the removal of \textit{:CurbYourEnthu} binding of \textit{?sitcom}.
In the end, $tp_1$ and $tp_3$ has the same set of triples,
but $tp_2$ now has only one triple \textit{(:Julia :actedIn :Seinfeld)}.
Now if we evaluate the original join $(tp_1 \leftouterjoin (tp_2 \Join tp_3))$
or a reordered one $(tp_1 \leftouterjoin tp_2) \leftouterjoin\ tp_3$
on these \textit{reduced} set of triples,
\textit{we do not need nullification to ensure consistent bindings of ?sitcom,
and there are no subsumed results}, because
each TP has a \textit{minimal} set of triples. \textbf{Minimality} of triples is defined as follows.

Let $\mathcal{R}$ be the final results of a query $Q$, and $tp$ be a TP in $Q$.
Let $tp.s$, $tp.p$, $tp.o$ be the respective subject, predicate, and object
positions in $tp$.
Let $Q$ be such that it SELECTs \textit{all} the variables as well as fixed positions
in \textit{all} the TPs in $Q$.
E.g., for the query in Figure \ref{fig:nullbest},
\textit{``SELECT :Jerry :hasFriend ?friend :actedIn ?sitcom :location :NewYorkCity WHERE...''}
selects \textit{everything} in the query, and the final results will be
\{\textit{(:Jerry, :hasFriend, :Larry, :actedIn, NULL, :location, :NewYorkCity),
(:Jerry, :hasFriend, :Julia, :actedIn, :Seinfeld, :location, :NewYorkCity)}\}. 
Note that this assumption of SELECTion
is just for the ease of definition of minimality, and not a required condition.
Let $\Delta_{tp}$ be the triples associated with a $tp$ after a semi-join or clustered-semi-join.
Then minimality of $\Delta_{tp}$ is defined as follows.
\begin{definition}
Let $\mathcal{R}_{tp} = $ (non-NULL)$\pi_{tp.s,tp.p,tp.o}(\mathcal{R})$, i.e., $\mathcal{R}_{tp}$
is a projection of the respective distinct bindings of $tp.s,tp.p,tp.o$ from $\mathcal{R}$ without any NULLs.
Then $\Delta_{tp}$ is said to be minimal, if $\Delta_{tp} = \mathcal{R}_{tp}$
($\Delta_{tp}$ and $R_{tp}$ can be empty as well).
In short, in a BGP-OPT query, the set of triples associated with a triple pattern
is minimal, if every triple creates one or more variable bindings
in the final results. There does not exist any triple which may get eliminated
as a result of an inner or left-outer-join.
\end{definition}

Next we see why minimality of triples is important in the context of an OPT pattern query.
\begin{thm}\label{lemma:min}
If every triple pattern in an OPT pattern has a minimal set of triples associated
with it, nullification and best-match operations are not required if an
original query $tp_1 \leftouterjoin (tp_2 \Join tp_3)$
is reordered as $(tp_1 \leftouterjoin tp_2) \leftouterjoin tp_3$. \qed
\end{thm}

The proof of Lemma \ref{lemma:min} is given in Appendix \ref{apdx:lemma:min}. 

Example-1 and Lemma \ref{lemma:min} together
interest us to find if the set of triples associated with each TP in an
OPT pattern can be reduced to minimal through semi-joins and clustered-semi-joins alone,
because then nullification and best-match can be avoided even if the joins are reordered.

\begin{figure}[h]
    \centering
        \includegraphics[width=3.4in, height=0.7in]{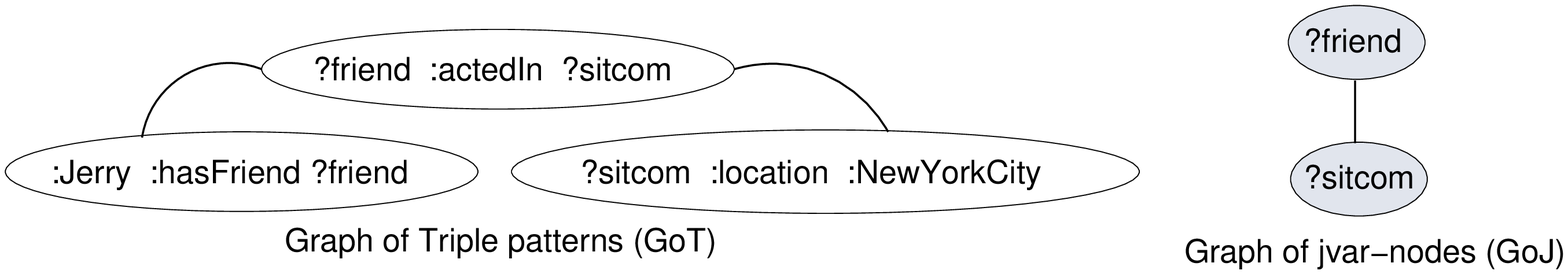}
       \caption{GoT and GoJ} \label{fig:got}
\end{figure}

\textbf{Acyclicity:}
Bernstein et al \cite{semij2,semij1} and Ullman \cite{ullman} have proved previously that if a ``graph
of tables'' (GoT) of an inner-join query is a ``tree'' (i.e., it is acyclic),
a bottom-up followed by a top-down pass
with semi-joins at each table in this tree, reduces the set of tuples in each table to a minimal.
In the context of a SPARQL query, GoT is a graph of TPs, where
each TP is treated as a unique table, and two TPs are connected with an undirected edge
if they share a join variable.
Any redundant cycles in GoT are removed as suggested in
\cite{semij2, semij1}\footnote{\scriptsize{Redundant cycles may occur if multiple TPs join over 
same variable.}}. 
We use and extend the acyclicity property of GoT for our optimization strategies.
For this, we construct a ``graph of join-variables'' (GoJ) as follows.
Each unique join variable (jvar) in a query is a node (jvar-node). There is an undirected edge
between two jvar-nodes, if they appear together in a TP.
Figure \ref{fig:got} shows GoT and GoJ for the query given in Figure \ref{fig:nullbest}.
For clarity and simplicity, we treat both GoT and GoJ separate from GoSN.
\begin{thm} \label{lemma:acyclic}
For a join query, if the GoT is acyclic, then the GoJ is acyclic too. \qed
\end{thm}
Acyclicity of GoJ follows from its construction.
We have given the proof of Lemma \ref{lemma:acyclic} in Appendix \ref{apdx:lemma:acyclic}. 
From Lemma \ref{lemma:acyclic}, we also observe the following property.
\begin{prop}
For a BGP query without OPT patterns (inner-joins only), if the GoJ is acyclic (is a tree),
a bottom-up followed by a top-down pass on the GoJ with ``clustered-semi-joins'' performed at each jvar-node,
reduces the set of triples associated with each TP in the query to a minimal. 
\end{prop} \label{prop:goj}

\textbf{Note:} We use the query and GoSN in Figure \ref{fig:qgraph}, its respective
data in Figure \ref{fig:nullbest}, and GoJ in Figure \ref{fig:got} together as a running example 
in the rest of the text in this paper without mentioning it explicitly every time. 

\subsection{Acyclic well-designed OPT patterns}\label{sec:semijloj}
In Section \ref{sec:bestm}, through Example-1 and Lemma \ref{lemma:min}, we showed
that if each TP in an OPT query has minimal triples, nullification and best-match
are not required. 
We showed how minimality of triples can be achieved for an acyclic OPT-free BGP
through Lemma \ref{lemma:acyclic} and Property \ref{prop:goj}.
\textit{But can we combine all these observations to use
the acyclicity of GoJ of a nested OPT query using its
graph of supernodes (GoSN)?} We explore it next.

Let the GoJ of a nested OPT pattern without Cartesian products
be acyclic, i.e., the GoJ is \textit{connected} and is a \textit{tree}.
Cyclic GoJ and Cartesian products are discussed separately
in Sections \ref{sec:cyclic} and \ref{sec:discuss} respectively. 
We choose the \textit{least selective} jvar node that appears
in an \textit{absolute master} supernode, and fix it as
the root of the GoJ tree. Jvar-nodes can be ranked for selectivity as follows:
A jvar-node $?j_1$ is considered more selective
than $?j_2$, if the most selective TP having $?j_1$ has fewer triples than
the most selective TP having $?j_2$, and so on.

With the root fixed as given before,
this GoJ tree has all the jvars in absolute masters towards the \textit{top} of
the tree (root and internal nodes), and jvars in the slaves towards the \textit{bottom}.
This observation follows from the facts that there are no Cartesian products, GoJ is
a tree, and we chose the \textit{root} from an absolute master.
We can traverse this tree bottom-up with semi-joins and clustered-semi-joins performed
at each jvar-node $?j$ as follows:
 
 $\bullet$ We perform a clustered-semi-join among the TPs that contain $?j$, and which appear
 either in the same supernode, or their supernodes are \textit{peers} of each other.
 
 $\bullet$ We perform a semi-join between pairs of TPs that are in a \textit{master-slave} relationship.

We can do a top-down pass following the same rules.
E.g., we choose \textit{?friend} as the root of GoJ tree
in our example, because it appears in absolute
master $SN_1$. In the bottom-up pass, we perform a clustered-semi-join
over \textit{?sitcom} among the peers $tp_2$
and $tp_3$, no semi-join for \textit{?sitcom}, because there
are no \textit{master-slave} TPs with \textit{?sitcom}.
We perform a semi-join $tp_2 \ltimes_{?friend} tp_1$,
no clustered-semi-join for \textit{?friend} because there are no peer TPs with \textit{?friend}.
Next, in the top-down pass, we process \textit{?friend} first followed by \textit{?sitcom}.
This process leaves each TP with a \textit{minimal} set of triples. 

\textit{But does this give us an optimal order of processing jvars?}
No, because a bottom-up pass
on the GoJ tree is same as processing OPT patterns per the order imposed
in the original query -- recall that all the jvars in
slaves appear towards the \textit{bottom} of the tree, and all the
jvars in masters towards the \textit{top} of the tree.
So this hardly fetches us any benefits of the selectivity of the master
TPs. \textit{We want to find an optimal order
of processing jvars, which exploits the selectivity of the masters
to aggressively prune the triples.}

For that we use \texttt{get\_jvar\_order} (Alg. \ref{alg:jvarord}).
From the GoJ tree of an OPT pattern, we consider an \textit{induced subtree} consisting only of
jvars that appear in the absolute master supernodes (ln \ref{ln:amtp}).
Since there are no Cartesian products, this induced subtree
is connected. We choose a jvar with the \textit{least selectivity} as the root
of this induced subtree, do a bottom-up pass on it, and store
the order in $order_{bu}$ (ln \ref{ln:root}--\ref{ln:bupass1}).
Choosing a jvar with least selectivity as the \textit{root} ensures that it is
processed \textit{last}. 

\begin{algorithm}[h]
\small{
\SetKw{Return}{return}
\SetKwFunction{bu}{bottom-up}
\SetKwFunction{td}{top-down}
\SetKwFunction{greedy}{greedy-jvar-order}
\SetKwInOut{Input}{input}
\SetKwInOut{Output}{output}
\Input{GoSN, GoJ}
\Output{$order_{bu}$, $order_{td}$}
\If{GoJ is cyclic}{
$order_{greedy}$ = \greedy{GoSN, GoJ}\;
\Return $order_{greedy}$, $order_{greedy}$ \label{ln:greedyord}
}
$\mathcal{J}_{m}$ = jvars in absolute master supernodes\;\label{ln:amtp}
$root$ = $?j \in \mathcal{J}_{m}$ with least selectivity\;\label{ln:root}
$\mathcal{T}_m$ = get-tree($\mathcal{J}_{m}$, $root$)\;
$order_{bu}$ = \bu{$\mathcal{T}_m$}\;\label{ln:bupass1}
$SN_{ss}$ = order remaining slaves with masters first\;\label{ln:orderslaves}
\BlankLine
\For{each slave supernode $SN_i$ in $SN_{ss}$}{\label{ln:slavetree}
  $\mathcal{J}_{s}$ = jvars in $SN_i$\;
  $root$ = $?j \in masters(SN_i)$\; \label{ln:mroot}
  $\mathcal{T}_s$ = get-tree($\mathcal{J}_{s}$, $root$)\;
  $order_{bu}$ = $order_{bu}$.append(\bu{$\mathcal{T}_s$})\;
}\label{ln:slavetreeend}
\BlankLine
$order_{td}$ = \td{$\mathcal{T}_m$}\;\label{ln:tdpass}
\For{each slave supernode $SN_i$ in $SN_{ss}$}{ \label{ln:slavetree2}
  $\mathcal{J}_{s}$ = jvars in $SN_i$\;
  $root$ = $?j \in masters(SN_i)$\;
  $\mathcal{T}_s$ = get-tree($\mathcal{J}_{s}$, $root$)\;
  $order_{td}$ = $order_{td}$.append(\td{$\mathcal{T}_s$})\;
} \label{ln:slavetreeend2}
\Return $order_{bu}, order_{td}$\;
}
\caption{\texttt{get\_jvar\_order}}\label{alg:jvarord}
\end{algorithm}
\vspace{-7mm}

\begin{algorithm}[h]
\small{
\SetKwFunction{CSJ}{clustered-semi-join}
\SetKwFunction{semij}{semi-join}
\SetKwFunction{slaveof}{slave-of}
\SetKwFunction{peerof}{peers-of}
\SetKwInOut{Input}{input}
\Input{$order_{bu}, order_{td}$, GoSN}
\For{each $?j$ in $order_{bu}$}{ \label{ln:prunebu}
  \For{each $tp_i$}{ \label{ln:busemi}
    \For{each $tp_j$}{
      \If{\slaveof{$tp_j$, $tp_i$}} {
	\semij($?j$, $tp_j$, $tp_i$)\tcp*{$tp_j \ltimes_{?j} tp_i$}
      }
    }
  } \label{ln:busemiend}
\BlankLine
  \For{each $SN_i$ with $?j$}{ \label{ln:bucsj}
    $S_{tp}$ = \{$tp\ |\ ?j \in tp, tp \in (SN_i \cup$  \peerof{$SN_i$})\}\;
    \CSJ{$?j$, $S_{tp}$}\;
  }\label{ln:bucsjend}
} \label{ln:prunebuend}
\BlankLine
\For{each $?j$ in $order_{td}$}{ \label{ln:tdpasssj}
  \For{each $tp_i$}{
    \For{each $tp_j$}{
      \If{\slaveof{$tp_j$, $tp_i$}} {
	\semij($?j$, $tp_j$, $tp_i$)\tcp*{$tp_j \ltimes_{?j} tp_i$}
      }
    }
  }
\BlankLine
  \For{each $SN_i$ with $?j$}{
  $S_{tp}$ = \{$tp\ |\ ?j \in tp, tp \in (SN_i \cup$ \peerof{$SN_i$})\}\;
  \CSJ{$?j$, $S_{tp}$}\;
  }
}\label{ln:tdpasssjend}
}
\caption{\texttt{prune\_triples}}\label{alg:prune}
\end{algorithm}

We order the remaining supernodes as -- masters before their respective slaves,
and among any two peer supernodes, a supernode with a more selective triple pattern
is ordered first. This order is $SN_{ss}$ (ln \ref{ln:orderslaves}).
Note that such an ordering of supernodes favours \textit{selective masters} to be processed before
their non-selective peers and slaves, thus benefiting the pruning process.
For each supernode $SN_i$ in $SN_{ss}$, we consider an induced subtree of GoJ
consisting only of jvars in $SN_i$, and choose a \textit{root} jvar
such that it also appears in a master of $SN_i$ (ln \ref{ln:mroot}).
Note that since GoJ is a \textit{connected tree}, a slave supernode
shares \textit{at least} one jvar with a master. 
We make a bottom-up pass on this induced subtree of $SN_i$, and
append it to $order_{bu}$ (ln \ref{ln:slavetree}--\ref{ln:slavetreeend}).
For a top-down pass, we reverse the above procedure. Starting with the induced subtree of
absolute masters, we do a top-down pass, and store it in $order_{td}$ (ln \ref{ln:tdpass}).
Using the same order of supernodes, $SN_{ss}$,
for an induced subtree of each $SN_i$ in $SN_{ss}$,
we do a top-down pass, and append it to $order_{td}$ (ln \ref{ln:slavetree2}--\ref{ln:slavetreeend2}).

With $order_{bu}$ and $order_{td}$ of jvar-nodes, we prune the
triples associated with TPs in a query using \texttt{prune\_triples} (Alg. \ref{alg:prune}).
For each jvar $?j$ in $order_{bu}$, first we do a \texttt{semi-join} between TPs that
are in a master-slave relationship (ln \ref{ln:busemi}--\ref{ln:busemiend}).
Then we do a \texttt{clustered-semi-join} among TPs having $?j$, and which appear in the
same supernode, or are peers of each other (ln \ref{ln:bucsj}--\ref{ln:bucsjend}).
Recall that the master-slave or peer relationship among TPs is determined by GoSN.
We repeat the same process, but now by following
$order_{td}$ of jvars (ln \ref{ln:tdpasssj}--\ref{ln:tdpasssjend}).
Notice that in this process,
\texttt{semi-joins} transfer the restrictions on variable bindings from master TPs  
to the slaves \textit{without} actually performing left-outer-joins,
and \texttt{clustered-semi-joins} transfer restrictions on variable bindings
among all the peers without actually doing inner-joins.
Through $order_{bu}$, $order_{td}$ of pruning we
ensure that jvars in masters always get pruned before those in slaves. 

\textbf{Example-2:}
Recalling our running example, which has an acyclic GoJ, 
we get $order_{bu}$ = [(\textit{?friend}), (\textit{?sitcom, ?fri-end})],
and $order_{td}$ = [(\textit{?friend}), (\textit{?friend, ?sitcom})] from
\texttt{get\_jvar\_order},
and with these we use \texttt{prune\_triples}
to do \texttt{semi-joins} and \texttt{clustered-semi-joins}
among the TPs in the master-slave and peer relationships respectively. 
\begin{thm} \label{lemma:optmin}
For a well-designed OPT query with an acy-clic GoJ, Algorithm \ref{alg:jvarord} followed by
Algorithm \ref{alg:prune} leaves a minimal set of triples for each triple pattern. \qed
\end{thm}
The proof of Lemma \ref{lemma:optmin} is given in Appendix \ref{apdx:lemma:optmin}. 

\subsection{Cyclic well-designed OPT patterns} \label{sec:cyclic}
For OPT-free BGPs, i.e., pure inner-joins, with cyclic GoJ,
minimality of triples cannot be guaranteed 
using clust-ered-semi-joins \cite{semij2,semij1,ullman}.
This result carries over immediately to cyclic OPT patterns too.
For a cyclic OPT pattern, minimality of triples cannot be guaranteed,
so we simply return $order_{greedy}$ from 
\texttt{get\_jvar\_order} (ln \ref{ln:greedyord}),
which is a \textit{greedy} order of jvars, i.e., all the jvars
are ranked in the descending order of their selectivity.
Recall from Section \ref{sec:semijloj},
that the relative selectivity between two jvars can be determined from the selectivity
of the TPs which have those jvars. In \texttt{prune\_triples},
we use $order_{greedy}$ in place of $order_{bu}$ and $order_{td}$, and follow
the rest of the procedure as is.
Since minimality of triples in each TP is not guaranteed, we need to use
the \textit{nullification} and \textit{best-match} operations in a reordered query
to ensure consistent variable bindings, and to remove any subsumed results.

This observation in general holds for all cyclic OPT queries, 
but we identify a \textit{subclass} of cyclic OPT queries
that can \textit{avoid} nullification and best-match by just using $order_{greedy}$ in place
of $order_{bu}$ and $order_{td}$ in \texttt{prune\_triples} --
\textit{in such cyclic OPT queries, each slave supernode has only one jvar in it}
(slaves can have one or more non-join variables). 
\begin{thm} \label{lemma:cyclenobestm}
For a well-designed OPT query with a cyclic GoJ, if each slave supernode has
only one jvar in it, nullification and best-match
can be avoided by using Algorithm \ref{alg:jvarord} followed by Algorithm \ref{alg:prune}. \qed
\end{thm}
We have given the proof in Appendix \ref{apdx:lemma:cyclenobestm}. 

From Alg. \ref{alg:jvarord} and \ref{alg:prune}, it may seem that \texttt{prune\_triples} is a 
``heavy-weight'' procedure, than simply doing pairwise joins between the TPs.
But as shown by our evaluation in  Section \ref{sec:eval}, our pruning procedure is
in fact quite ``light-weight'', especially for low-selectivity complex OPT patterns
(please note the $T_{prune}$ values compared to $T_{total}$
in the query times, and ``\#initial triples'' and ``\#triples aft pruning'' columns
in Tables \ref{tbl:eval1}, \ref{tbl:eval2}, \ref{tbl:eval3}).
We achieve this through the usage of compressed bitvector indexes on RDF graphs,
and procedures that directly work on these compressed indexes without decompressing them.
These are described in Section \ref{sec:bitmat}. Together they give a competitive query performance.

\section{Indexing the RDF graph} \label{sec:bitmat}
In the recent few years, there have been a lot of advances in the efficient storage and indexing of
RDF graphs, with many research systems, e.g., RDF-3X \cite{rdf3x}, TripleBit \cite{triplebit}, BitMat \cite{bitmatwww10},
as well as large scale open-source and commercial engines, e.g., Virtuoso \cite{virtuoso},
MonetDB \cite{monetdb, swans}.
Among these options, we have chosen BitMats as the base index structure for implementing
our technique.
Our reasons of choice are given after we give a brief overview of BitMat,
our enhancements in it,
and the \texttt{fold, unfold} operations for the completeness of the text.

\begin{figure}[h]
    \centering
	\includegraphics[scale=0.3]{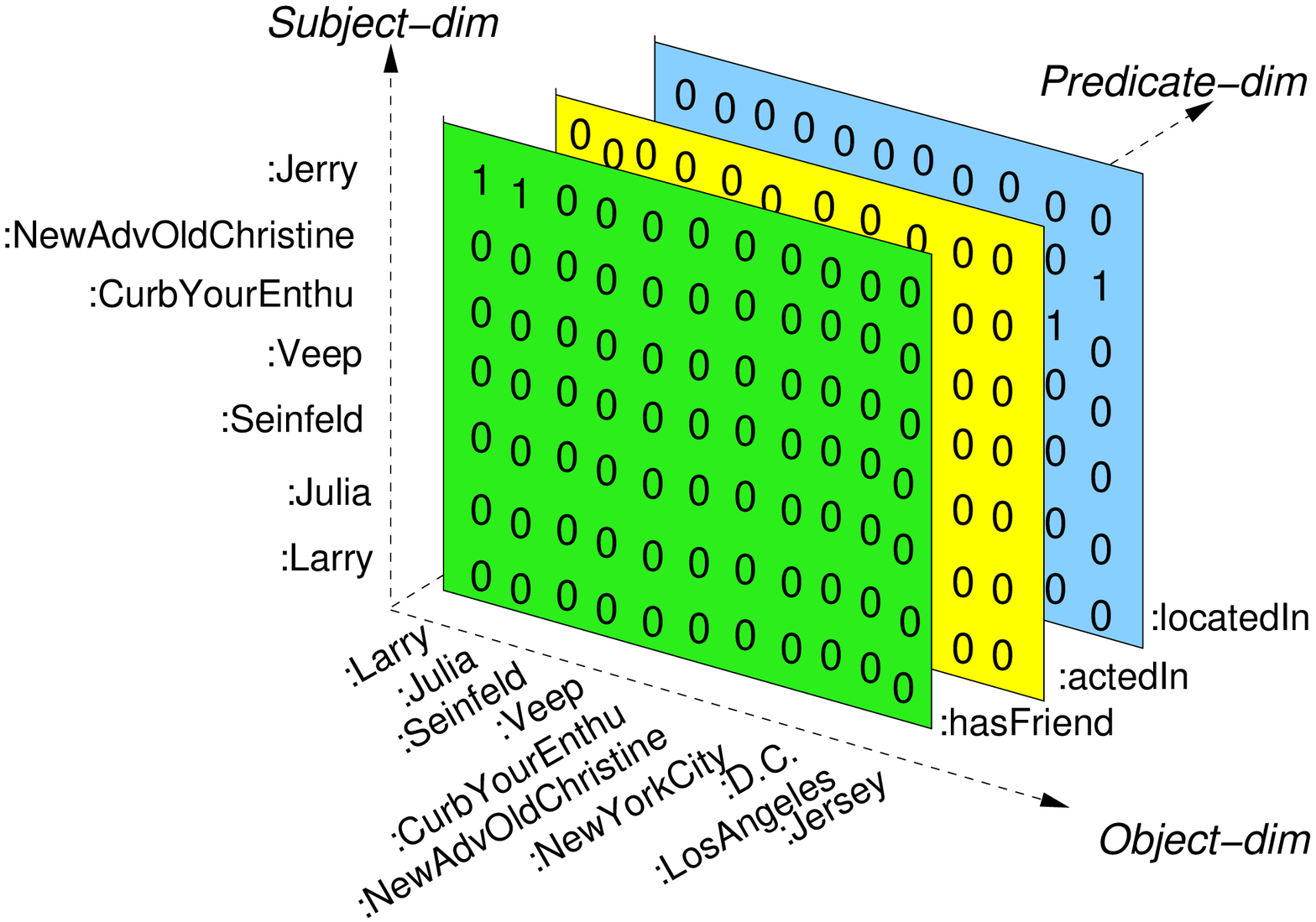}
        \caption{3D Bitcube of RDF data in Figure \ref{fig:nullbest}}
    \label{fig:bitcube}
\end{figure}

If $V_s$, $V_p$, and $V_o$ are
the sets of unique subject, predicate, and object values in an RDF dataset,
then a 3D bitcube of RDF data has $V_s \times V_p \times V_o$ dimensions.
Each cell in this bitcube represents a unique RDF triple formed by the coordinate values (S P O).
If this (S P O) triple is present in the given RDF dataset, that bit is set to 1 in the bitcube.
To facilitate joins on S-O dimensions, same 
S and O values are mapped to the same coordinates of the respective
dimensions\footnote{\scriptsize{For the scope of this paper, we do not consider joins on S-P
or O-P dimensions.}}.
Due to space constraints, the exact details of this procedure
are given in Appendix \ref{apdx:bitmat} for reader's convenience.
They are borrowed from the details given in Section 3 of \cite{bitmatwww10}.
The 3D bitcube of the data given in Figure \ref{fig:nullbest} is shown in Figure \ref{fig:bitcube}.

This bitcube is conceptually sliced along each dimension, and the 2D BitMats are created.
In general, four types of 2D BitMats are created: (1) S-O and O-S BitMats by slicing
the P-dimension (O-S BitMats are nothing but \textit{transpose} of the respective S-O BitMats),
(2) P-O BitMats by slicing the S-dimension, and (3) P-S BitMats by slicing the O-dimension.
Altogether we store $2*|V_p| + |V_s| + |V_o|$ BitMats for any RDF data.
Figure \ref{fig:bitcube} shows 2D S-O BitMats that we can get by slicing the predicate dimension
(others are not shown for conciseness).
Intuitively, a 2D S-O or O-S BitMat of predicate \textit{:hasFriend}
represents all the triples matching a triple pattern of kind (\textit{?a :hasFriend ?b}), 
a 2D P-S BitMat of O-value \textit{:Seinfeld} represents all triples matching triple pattern
(\textit{?c ?d :Seinfeld}), and so on.

Each row of these BitMats is compressed using run-length-encoding.
A bit-row like ``1110011110'' is represented as ``[1] 3 2 4 1'', and  
``0010010000'' is represented as ``[0] 2 1 2 1 4''. Notably, in the second case,
the bit-row has only two set bits, but it has to use five integers in the
compressed representation. So we use a \textit{hybrid} representation in our
implementation that works as follows -- if the number of set bits in a bit-row are less than 
the number of integers used to represent it, then we simply store the set bit positions.
So ``0010010000'' will be compressed as ``3 6'' (3 and 6 being
the positions of the set bits). This hybrid compression fetches us as much as
40\% reduction in the index space compared to using only run-length-encoding as done in \cite{bitmatsrc}.

\textbf{\textit{Fold}} operation is represented as `\texttt{fold(BitMat, RetainDimension) returns bitArray}'.
It takes a 2D BitMat and \textit{folds} it by retaining the \textit{RetainDimension}.
More succinctly, a fold operation is nothing but \textit{projection} of distinct values of the particular
BitMat dimension, by doing a bitwise OR on the other dimension. It can be represented as:
\vspace{-2mm}
\[
\mathtt{fold(BM_{tp}, dim_{?j})} \equiv \pi_{?j}(BM_{tp})
\]

\vspace{-2mm}
$BM_{tp}$ is a 2D BitMat holding the triples matching $tp$,
and $dim_{?j}$ is the dimension of BitMat that represents variable $?j$ in $tp$.
E.g., for a triple pattern (\textit{?friend :actedIn ?sitcom}), if we consider the O-S BitMat of predicate
\textit{:actedIn}, \textit{?friend} values are in the ``column'' dimension, and \textit{?sitcom} values
are in the ``row'' dimension of the BitMat.

\textbf{\textit{Unfold}} is represented as `\texttt{unfold(BitMat, MaskBitArray,
RetainDimension)}'. For every bit set to 0 in the \textit{MaskBitArray}, \textit{unfold}
clears all the bits corresponding to that position of the \textit{RetainDimension}
of the BitMat. \textit{Unfold} can be simply represented as:
\vspace{-2mm}
\[
\mathtt{unfold(BM_{tp}, \beta_{?j}, dim_{?j})} \equiv \{t\ |\ t \in BM_{tp}, t.?j \in \beta_{?j}\}
\]

\vspace{-2mm}
$t$ is a triple in $BM_{tp}$ that matches $tp$.
$\beta_{?j}$ is the \textit{MaskBitArray} containing bindings of $?j$ to be retained.
$dim_{?j}$ is the dimension of $BM_{tp}$ that represents $?j$, and $t.?j$ is a binding
of $?j$ in triple $t$. In short, \texttt{unfold} keeps only those triples whose respective
bindings of $?j$ are set to 1 in $\beta_{?j}$, and removes all other.

\textbf{Reasons for choice of BitMats:}
RDF stores that achie-ve high compression ratio using
variable length or \textit{delta} encoding, have to decode/decompress the
\textit{variable length} IDs, and get them in 4-byte integers for performing joins.
This ends up being an overhead for queries that have to access a large amount of data.
In BitMats, IDs and in turn triples are represented by \textit{bits} compressed with 4-byte run-lengths,
which are manipulated through the \texttt{fold-unfold} procedures without decompressing them.
Hence we have chosen BitMats as the base index structure in our implementation.

\section{Query Processing} \label{sec:qproc}
Up to this point, we saw how to construct
a GoSN to capture the nesting of OPT patterns in a query
in Section \ref{sec:qgrconstr1}. In Sections \ref{sec:semijloj} and \ref{sec:cyclic},
we presented our optimization strategies 
to ``prune'' the triples associated with TPs in acyclic and cyclic well-designed queries,
without considering the underlying index structure for RDF data.
Then in Section \ref{sec:bitmat}, we described BitMat -- our index structure. 

Now, in this section, we ``connect the dots'', i.e., we present how triples
associated with TPs in a query are pruned using BitMats, 
how \texttt{semi-join} and \texttt{clustered-semi-join} in
\texttt{prune\_triples} (Alg \ref{alg:prune})
are achieved through BitMats and \texttt{fold-unfold} by consulting GoSN,
and finally how output results are generated using a
\textit{multi-way pipelined join} for any \textit{acyclic or cyclic well-designed} query.
We put all these concepts together in Algorithm \ref{alg:qproc}. 

\begin{algorithm}[h]
\small{
\SetKwFunction{getgosn}{get-graph-supernodes}
\SetKwFunction{getgoj}{get-graph-jvars}
\SetKwFunction{getjord}{get\_jvar\_order}
\SetKwFunction{bestmatch}{decide-best-match-reqd}
\SetKwFunction{init}{init}
\SetKwFunction{prune}{prune\_triples}
\SetKwFunction{bestmop}{best-match}
\SetKwFunction{finalres}{multi-way-join}
\SetKw{return}{return}
\SetKwInOut{Input}{input}
\SetKwInOut{Output}{output}
\Input{Original BGP-OPT query}
\Output{Final results}
\BlankLine
GoSN = \getgosn{Orig BGP-OPT query}\; \label{ln:getgosn}
GoJ = \getgoj{Orig BGP-OPT query}\; \label{ln:getgoj}
\For{each $tp_i$ in GoSN}{
\{$tp_i \Rightarrow BM_{tp_i}$\} = \init{}\; \label{ln:init}
}
\BlankLine
\textbf{bool} \textit{NB-reqd} = \bestmatch{GoSN, GoJ}\; \label{ln:bestmdec}
\tcp{Alg \ref{alg:jvarord}}
($order_{bu}, order_{td}$) = \getjord{GoSN, GoJ}\label{ln:callgetjord}
\BlankLine
\prune{$order_{bu}$, $order_{td}$, GoSN}\tcp*{Alg \ref{alg:prune}}\label{ln:callprune}
\BlankLine
sorted-tps = sort TPs in master-slave hierarchy\; \label{ln:sorttps}
\tcp{Final result generation - Alg \ref{alg:finalres}} 
\textit{allres} = \finalres{vmap, sorted-tps, visited, NB-reqd}\; \label{ln:finalres}
\BlankLine
\eIf{NB-reqd}{ 
  \textit{finalres} = \bestmop{allres}\; \label{ln:bestmatch}
}{
  \textit{finalres} = \textit{allres};
}
\return{finalres}\;
}
\caption{Query processing}\label{alg:qproc}
\end{algorithm}
\vspace{-7mm}

\begin{algorithm}[h]
\small{
\SetKwFunction{fold}{fold}
\SetKwFunction{unfold}{unfold}
\SetKw{return}{return}
\SetKwInOut{Input}{input}
\Input{$?j$, $tp_j$, $tp_i$}
\BlankLine
$\beta_?j$ = \fold{$BM_{tp_i}$, $dim_{?j}$} AND \fold{$BM_{tp_j}$, $dim_{?j}$}\; \label{ln:collect}
\unfold{$BM_{tp_j}$, $\beta_{?j}$, $dim_{?j}$}\; \label{ln:distr}
}
\caption{\texttt{semi-join}}\label{alg:semij}
\end{algorithm}
\vspace{-7mm}

\begin{algorithm}[h]
\small{
\SetKwFunction{fold}{fold}
\SetKwFunction{unfold}{unfold}
\SetKw{return}{return}
\SetKwInOut{Input}{input}
\Input{$?j$, \{$tp_1, ..., tp_k$\}}
\BlankLine
$\beta_{?j}$ = bitarray with all bits set to 1\;
\For{each $tp_i$ in \{$tp_1, ..., tp_k$\}}{
$\beta_{?j}$ = $\beta_{?j}$ AND \fold{$BM_{tp_i}$, $dim_{?j}$}\;
}
\For{each $tp_i$ in \{$tp_1, ..., tp_k$\}}{
\unfold{$BM_{tp_i}$, $\beta_{?j}$, $dim_{?j}$}
}
}
\caption{\texttt{clustered-semi-join}}\label{alg:csemij}
\end{algorithm}

In Algorithm \ref{alg:qproc}, we first construct the GoSN and GoJ (ln \ref{ln:getgosn}-\ref{ln:getgoj}).
Then with \texttt{init}, we load a BitMat for each TP in the query
that contains the triples matching that TP (ln \ref{ln:init}).
Recall from Section \ref{sec:bitmat}, that we have, in all, four types of BitMats.
We choose an appropriate BitMat for each TP as follows.
If the TP in the query is of type (?var :fx1 :fx2),
i.e., with two fixed positions, we load only one row corresponding to :fx1
from the P-S BitMat for :fx2. Similarly for a TP of type (:fx1 :fx2 ?var),
we load only one row corresponding to :fx2 from the P-O BitMat for :fx1.
E.g., for (\textit{?sitcom :location :NewYorkCity}) we load only one
row corresponding to \textit{:location} from the P-S BitMat of \textit{:NewYorkCity}.
If the TP is of type (?var1 :fx1 ?var2), we load either the
S-O or O-S BitMat of :fx1. If ?var1 is a join variable
and ?var2 is not, we load the S-O BitMat and vice versa. If both, ?var1 and ?var2,
are join variables, then we check which of ?var1 and ?var2 appears in $order_{bu}$
before the other. If ?var1 comes before ?var2, we load the S-O BitMat and vice versa.
Recalling Example-2 from Section \ref{sec:semijloj},
for (\textit{?friend :actedIn ?sitcom}) we load the S-O BitMat of
\textit{:actedIn} because \textit{?friend} comes before \textit{?sitcom} in $order_{bu}$. 

While loading the BitMats with \texttt{init}, we do active pruning
using the TPs that may have been initialized previously.
E.g., if we first load BitMat $BM_{tp_1}$ containing triples matching
(\textit{:Jerry :hasFriend ?friend}), then 
while loading $BM_{tp_2}$, we use the bindings of \textit{?friend}
in $BM_{tp_1}$ to actively prune the triples
in $BM_{tp_2}$ while loading it. Then while loading $BM_{tp_3}$,
we use the bindings of \textit{?sitcom} in $BM_{tp_2}$ to actively prune the triples
in $BM_{tp_3}$.
We check whether two TPs are joining with each other over an inner or left-outer join
using GoSN with the \textit{master-slave} or \textit{peer} relationship,
and then decide whether to use other BitMat's variable bindings.

Next we decide if nullification and best-match are required -- they are
required for a cyclic query where slaves have more than one jvars (ln \ref{ln:bestmdec}).
Then using \texttt{get\_jvar\_order} (ln \ref{ln:callgetjord}) we get
an optimal order of jvars for the bottom-up and top-down passes
on GoJ. Recall that for a \textit{cyclic} GoJ, \texttt{get\_jvar\_order}
simply returns ($order_{greedy}$, $order_{greedy}$).
Next, we prune the triples in BitMats using \texttt{prune\_triples}
(ln \ref{ln:callprune} in Alg \ref{alg:qproc}). This procedure
has already been explained in Section \ref{sec:semijloj}, hence we refer the reader
to it. Two important operations in \texttt{prune\_triples} are \texttt{semi-join} and \texttt{clustered-semi-join}
-- to remove the triples from the BitMats. These operations
make use of the \texttt{fold} and \texttt{unfold} primitives.
We have shown how \texttt{fold} and \texttt{unfold} are used in
\texttt{semi-join} and \texttt{clustered-semi-join} with 
Algorithms \ref{alg:semij} and \ref{alg:csemij} respectively.

Recall from \texttt{prune\_triples} (Alg \ref{alg:prune})
that we do a \texttt{semi-join} (Alg \ref{alg:semij}) between two TPs,
if they are in a master-slave relationship over a shared a jvar.
The slave TP takes the restrictions on the variable bindings
from the master. In \texttt{fold} we project out the bindings of
$?j$ from $tp_i$ and $tp_j$ in the form of 
bitarrays. Through a bitwise AND we take an intersection of these bindings
and store the intersection result in $\beta_{?j}$.
Then we use $\beta_{?j}$ as the \textit{MaskBitArray} in \texttt{unfold}
to remove any triples whose respective bindings for $?j$ were dropped
as a result of the intersection.
\texttt{Clustered-semi-join} (Alg \ref{alg:csemij}) is same as \texttt{semi-join}, except that
we transfer the restrictions on the variable bindings
across all the TPs that share a join variable and are \textit{peers} of each other.
Recall Example-2 from Section \ref{sec:semijloj}, which shows how \texttt{semi-join}
and \texttt{clustered-semi-join} are used.

If an OPT query is \textit{acyclic}, after \texttt{prune\_triples},
each TP BitMat has a \textit{minimal} set of triples (Lemma \ref{lemma:optmin}).
In case of a \textit{cyclic} OPT query, \texttt{prune\_triples} 
only reduces the triples in the BitMats, but they may not be minimal.
Note that using \texttt{prune\_triples}, we prune the triples
in BitMats, but we need to actually ``join'' them to produce the final results.
For that we use \texttt{multi-way-join} (ln \ref{ln:finalres} in Alg \ref{alg:qproc}).
This procedure is described separately in Section \ref{sec:finalres}.
After \texttt{multi-way-join}, we use \texttt{best-match} to remove any subsumed results
only if the query is cyclic and its slaves have more than one jvars 
(ln \ref{ln:bestmatch}) -- recall Lemmas \ref{lemma:optmin} and \ref{lemma:cyclenobestm}.
In \texttt{best-match}, we externally sort all the results
generated by \texttt{multi-way-join}, and then 
remove the subsumed results with a single pass over them.

In the \texttt{init} and \texttt{prune\_triples} processes,
we do a ``simple optimization'' -- if at any point, a TP in an absolute
master supernode has zero triples, we take that as a hint of an empty result,
and abandon any further query processing.

Currently Algorithm \ref{alg:qproc} is a main-memory process,
i.e., all the TP BitMats are kept in memory during the query processing,
and there is no disk spooling.
This may pose some limitations on the total size of BitMats in a query.
But as seen in our evaluation, LBR could handle low-selectivity queries with up
to 13 TPs on a dataset with more than a billion triples on a machine with 8 GB memory,
exhibiting the scalability of our technique.
Presently LBR does not handle TPs with all variable positions (\textit{?a ?b ?c}), and supporting
them is currently under development.

\subsection{Multi-way Pipelined Join} \label{sec:finalres}
Before calling \texttt{multi-way-join}, we first sort all the TPs in the query as follows.
Considering the TPs in absolute master supernodes, we sort them in an ascending
order of the number of triples left in each TP's BitMat.
Then we sort remaining TPs in the descending order of master-slave hierarchy and selectivity.
That is, among two supernodes connected as $SN_1 \rightarrow SN_2$,
TPs in $SN_1$ and any \textit{peers} 
of $SN_1$ are sorted before those in $SN_2$. Among the peer TPs, they are
sorted in the ascending order of the number of triples left in their BitMats
(ln \ref{ln:sorttps} in Alg \ref{alg:qproc}).
This order is \texttt{stps}.
In \texttt{multi-way-join} we use at most $\sum_{tp_i \in Q} vars(tp_i)$
additional memory buffer, where $vars(tp_i)$ are the
variables in every $tp_i$ in the query $Q$.
This is \texttt{vmap} in Alg \ref{alg:finalres}.
Thus we use negligible additional memory in \texttt{multi-way-join}.

At the beginning, \texttt{multi-way-join} gets an empty \texttt{vmap} for storing the 
variable bindings, \texttt{stps},
an empty \texttt{visited} list, and a flag \texttt{nulreqd} indicating if nullification
is required (depending on the cyclicity of the query).
In \texttt{multi-way-join},
we go over each triple in $BM_{tp_1}$ of the first TP in \texttt{stps},
generate bindings for the variables in $tp_1$, and
store them in \texttt{vmap}. We add $tp_1$ to the \texttt{visited} list,
and call \texttt{multi-way-join} 
recursively for the rest of the TPs (ln \ref{ln:firsttp}--\ref{ln:endfirsttp}).
Note that in each recursive call, \texttt{multi-way-join} gets a partially
populated \texttt{vmap} and a \texttt{visited} list that tells which TP's
variable bindings are already stored in \texttt{vmap}.
Then we check if any variables in a non-visited $tp_i$ are already mapped
in \texttt{vmap} (ln \ref{ln:getbind}--\ref{ln:continue}).
Recall that since the query does not have Cartesian products, we always find
\textit{at least} one $tp_i$, which has one or more of its variables
mapped in \texttt{vmap}.
Also notice that \texttt{stps} order ensures that a master TP's variable
bindings are stored in \texttt{vmap} before its slaves. 
If there exist one or more triples $t$ in $BM_{tp_i}$ consistent with the variable bindings
in \texttt{vmap}, then for each such $t$ we generate bindings 
for all the variables in $tp_i$, store them in \texttt{vmap}, and proceed with the recursive call
to \texttt{multi-way-join} for the rest of the TPs (ln \ref{ln:mapother}--\ref{ln:endmapother}).
Notice that, this way we \textit{pipeline} all the BitMats, and do not do pairwise joins or use any
other intermediate storage like hash-tables. 

If we do not find any triple in $BM_{tp_i}$ consistent with the existing variable bindings in \texttt{vmap},
then -- (1) if $tp_i$ is an absolute master, we \textit{rollback} from this point, because an absolute master TP
cannot have NULL bindings (ln \ref{ln:rollback}), else (2) we map all the variables
in $tp_i$ to NULLs, and proceed with the recursive call to \texttt{multi-way-join} (ln \ref{ln:nomap}--\ref{ln:endnomap}).
When all the TPs in the query are in the \texttt{visited} list, we check if we require \texttt{nullification}
to ensure consistent variable bindings in \texttt{vmap} across all the slave TPs, and \texttt{output} one result
(ln \ref{ln:outputres}--\ref{ln:endoutputres}).
We continue this recursive procedure till triples in $BM_{tp_1}$ are exhausted
(ln \ref{ln:firsttp}--\ref{ln:endfirsttp}).

\begin{algorithm}[h]
\small{
\SetKw{return}{return}
\SetKw{continue}{continue}
\SetKw{true}{true}
\SetKw{false}{false}
\SetKwFunction{finalres}{multi-way-join}
\SetKwFunction{output}{output}
\SetKwFunction{masterof}{master-of}
\SetKwFunction{peerof}{peer-of}
\SetKwFunction{nuli}{nullification}
\SetKwInOut{Input}{input}
\SetKwInOut{Output}{output}
\Input{vmap, stps, visited, nulreqd}
\Output{all the results of the query}
\BlankLine
\If{visited.size == stps.size}{\label{ln:outputres}
  \If{nulreqd}{
    \nuli(vmap)\; \label{ln:nullification}
  }
  \output(vmap)\tcp*{generate a single result} \label{ln:endoutputres}
\return\;
}
\eIf{visited is empty}{ \label{ln:firsttp}
  $tp_1$ = first TP from stps\;
  visited.add($tp_1$)\;
  \For{each triple $t \in BM_{tp_1}$}{
    generate bindings for vars($tp_1$) from $t$, store in vmap\;
    \finalres{vmap, stps, visited, nulreqd}\;
  } \label{ln:endfirsttp}
}{
  \For{each $tp_i$ in stps}{
    \If{$tp_i \in$ visited}{
      \continue\;
    }
     get bindings for vars($tp_i$) from vmap\; \label{ln:getbind}
     \If{no bindings found}{
      \continue\;
     } \label{ln:continue}
\BlankLine
    \textit{atleast-one-triple} = \false\;
     \For{each triple $t \in BM_{tp_i}$ with same bindings} {\label{ln:mapother}
      \textit{atleast-one-triple} = \true\;
      store vars($tp_i$) bindings from $t$ in vmap\;
      visited.add($tp_i$)\;
      \finalres{vmap, stps, visited, nulreqd}\;
      visited.remove($tp_i$)\; \label{ln:endmapother}
    }
\BlankLine
  \If{(\textit{atleast-one-triple} == \false)}{
      \If{$tp_i$ is an absolute master}{
      \return\; \label{ln:rollback}
      }
     \tcp{This means $tp_i$ is a slave}
      set all vars($tp_i$) to NULL in vmap\; \label{ln:nomap}
      visited.add($tp_i$)\;
      \finalres{vmap, stps, visited, nulreqd}\;
      visited.remove($tp_i$)\;\label{ln:endnomap}
  }
 }
}
}
\caption{\texttt{multi-way-join}}\label{alg:finalres}
\end{algorithm}

Intuitively, \texttt{multi-way-join} is reminiscent of a relational join plan
with reordered left-outer-joins -- that is, in \texttt{stps} we sort
selective masters before their non-selective peers and slaves, and
masters generate variable bindings before slaves in \texttt{vmap}.
\textit{But note that we pipeline all the joins together, and
we can skip nullification and best-match
for acyclic queries and cyclic queries with only one jvar per slave,
because of our optimization techniques},
\texttt{get\_jvar\_order} (Alg \ref{alg:jvarord}), \texttt{prune-\_triples}
(Alg \ref{alg:prune}), and Lemmas \ref{lemma:optmin} and \ref{lemma:cyclenobestm}. 

Recall Example-2 from Section \ref{sec:semijloj}.
After \texttt{prune\_triples}, $tp_1$ has
two triples, and $tp_2$, $tp_3$ have one triple each in their BitMats.
We sort the TPs as \texttt{stps} = [$tp_1$, $tp_2$, $tp_3$]. 
In \texttt{multi-way-join}, we first generate a binding of $tp_1.?friend$
and store it in \texttt{vmap}. In the recursive calls,
we locate triples with the same \textit{?friend} binding
in $BM_{tp_2}$. For each such triple, we generate $tp_2.?friend$,
$tp_2.?sitcom$ bindings in \texttt{vmap}, and proceed to $tp_3$.
In a recursive call, if we do not find any triple
with the same variable bindings in $tp_2$ or $tp_3$, we set variables
in that TP to null in \texttt{vmap}.
Since this is an acyclic query, we do not need \texttt{nullification}.
While \texttt{output}ting a result, we pick variable
bindings generated by masters over their slaves
for common variables in \texttt{vmap}, e.g.,
we pick binding of \textit{?friend} from $tp_1.?friend$ over $tp_2.?friend$. 

\subsection{Discussion} \label{sec:discuss}
For our experiments, we assume that \textit{all} the variables in a query -- join as well as 
non-join -- are SELECTed for projection, because analysis of DBPedia query logs shows that
over 95\% of the queries SELECT all the variables \cite{swim}.
As per W3C specifications, SPARQL algebra follows ``bag semantics''
\cite{sparql}, so SELECTion (projection) of particular variables from a query
can be readily supported in LBR by just intercepting the \texttt{output(vmap)}
statement at line-\ref{ln:endoutputres} in \texttt{multi-way-join}, which 
will output only mappings of the SELECTed variables from \texttt{vmap}. 

For the scope of this paper, we have focused on the ``join'' component of SPARQL,
i.e., BGP-OPT patterns without UNIONs, FILTERs, or Cartesian products.
But for the completeness of the text, here 
we discuss how our technique can be extended to handle them.

\textbf{UNION:}
For this discussion, we assume \textit{well-designed} UNIONs (UWD), which are --
for every subpattern $P'=(P_1 \cup P_2)$ in a query,
if a variable $?j$ in $P'$ appears outside $P'$, then it appears in both $P_1$ and $P_2$.
UWDs tend to have high occurrence (e.g., 99.97\% as shown in \cite{swim}).
Also UWDs remain unaffected by the difference between SPARQL and SQL over
the treatment of NULLs \cite{polleres}, and
following equivalences hold on them \cite{perez2} --
(1) $(P_1 \cup P_2) \leftouterjoin P_3 \equiv (P_1 \leftouterjoin P_3) \cup (P_2 \leftouterjoin P_3)$,
(2) $(P_1 \cup P_2) \Join P_3 \equiv (P_1 \Join P_3) \cup (P_2 \Join P_3)$.
Additionally we also rewrite - (3) $P_1 \leftouterjoin (P_2 \cup P_3)$ as 
$(P_1 \leftouterjoin P_2) \cup (P_1 \leftouterjoin P_3)$,
and remove any spurious results in the end\footnote{\scriptsize{Spurious results may get introduced
if either $P_2$ or $P_3$ is empty,
or $P_2$ or $P_3$ does not have any matching variable bindings to that of $P_1$.}}.
Using these rules, we rewrite a \textit{well-designed} BGP-OPT-UNION query  
in the \textit{UNION normal form} (UNF) \cite{perez2}, i.e., a query of
the form $P_1 \cup P_2 \cup\ ..\ \cup P_n$ where each sub-pattern $P_i$ ($1 \leq i \leq n$) is UNION-free.
We evaluate each UNION-free $P_i$ using the same LBR technique,
remove any spurious results if rule (3) of rewrite is used, and
\textit{add} the results from all the $P_i$s.
Note that we can \textit{add} the output of all the $P_i$s without a conventional \textit{set-union},
because SPARQL UNION follows ``bag semantics'' \cite{sparql,schmidt}. 

\textbf{FILTER:} For this discussion, we assume \textit{safe} FILTERs \cite{perez2,polleres},
i.e., for $P_x \mathcal{F}(R)$, where $R$ is a filter to be applied
on $P_x$, all the variables in $R$ appear in $P_x$, $vars(R) \subseteq vars(P_x)$.
\textit{Unsafe} filters can alter the semantics of OPT patterns (ref \cite{perez2}).
In addition to the previous rewrite rules introduced under UNION,
for a \textit{well-designed} BGP-OPT-UNION query with \textit{safe} filters, following equivalences hold
\cite{perez2, schmidt} --
(4) $(P_1 \leftouterjoin P_2)\mathcal{F}(R) \equiv (P_1 \mathcal{F}(R)) \leftouterjoin P_2$,
(5) $(P_1 \cup P_2)\mathcal{F}(R) \equiv (P_1 \mathcal{F}(R)) \cup (P_2 \mathcal{F}(R))$.
Using these five rewrite rules, we \textit{push in} filters,
\textit{push out} unions, and bring the query in the UNF,
i.e., $P_1 \cup P_2 \cup...\cup P_n$ where each sub-pattern $P_i$ ($1 \leq i \leq n$) is union-free.

Now we can evaluate each $P_i$ using an augmented LBR technique as follows.
We apply filters either by intercepting \texttt{init} (ln \ref{ln:init}) and
\texttt{prune\_triples} (ln \ref{ln:callprune}) in Alg \ref{alg:qproc}, or 
we apply them using a new \textit{filter-and-nullification} (\texttt{FaN}) routine,
in place of \texttt{nullification} (ln \ref{ln:nullification} in \texttt{multi-way-join}).
The decision depends on the type of filters, e.g., if a filter has
variables from two or more  TPs, then we 
apply it in \texttt{FaN} when variable bindings from all the TPs are generated in
\texttt{vmap}. Through \texttt{FaN}, we apply filters as well as do nullification for a cyclic $P_i$.
We apply \texttt{best-match} in the end if \texttt{FaN} nullifies \textit{at least}
one variable binding in \texttt{vmap} or $P_i$ is cyclic.
We remove any spurious results introduced due to rule (3)
of UNION rewrite, and obtain the final results by adding the
results from each $P_i$ in the UNF --
please recall our note about SPARQL ``bag semantics'' under UNION.
We can use some ``cheap'' filter optimizations before rewriting a query,
e.g., a filter $P_1 \mathcal{F}(?m = ?n)$ can be eliminated by
replacing every $?n$ by $?m$ in all the TPs in $P_1$.
Extending LBR technique for OPT patterns with
UNIONs and FILTERs is a part of our ongoing work.

\textbf{Cartesian products ($\times$):} 
If a query is in the \textit{Cartesian normal form}, i.e., 
$P_1 \times P_2 \times .. \times P_n$ where each sub-pattern $P_i$ ($1 \leq i \leq n$) is $\times$-free,
then we can evaluate each $\times$-free $P_i$ using the same LBR technique presented in this paper,
and generate the final results by taking a cross product of the results from each $P_i$ sub-pattern.
However, a query with an arbitrary nesting of $\leftouterjoin$ and $\times$ poses challenges,
because $\leftouterjoin$ is \textit{not distributive} over $\times$.
A na\"{\i}ve way of handling such a query can be -- use the technique
presented in this paper only for the sub-patterns that are $\times$-free, and then resort to the
standard relational technique of performing pairwise joins or $\times$-products to get the final results.
As a part of the future work, we plan to extend LBR's technique
to handle such nested OPT-Cartesian queries using an \textit{augmented} GoSN construction
and modified \texttt{multi-way-join} (Alg \ref{alg:finalres}).
Note that queries with Cartesian products are rare for RDF data.
Previously published SPARQL queries in the literature
do not have Cartesian products \cite{bitmatwww10,triad,dbpediaspbench,rdf3x,swim,triplebit,gstore}. 

\section{Evaluation} \label{sec:eval}
We developed \textit{Left Bit Right} (LBR) query processing technique with C/C++ language,
compiled with g++ v4.8.2, -O3 -m64 flags, on a 64 bit Linux
3.13.0-34-generic SMP kernel (Ubuntu 14.04 LTS distribution).
For running the experiments, we used a Lenovo T540p laptop with
Intel Core i3-4000M 2.40GHz CPU, 8 GB memory, 12 GB swap space, and 1 TB Western Digital
5400RPM SATA hard disk.

\subsection{Setup for Experiments}
\textbf{RDF Stores:}
To evaluate the competitive performance of LBR's techniques, we used two columnstores
Virtuoso v.7.1.0 \cite{virtuoso} and MonetDB v11.17.21 \cite{monetdb}, which are popular
for RDF data storage.
LBR's core query processing algorithm works with the integer IDs assigned
to the subjects, predicates, objects.
Hence we loaded RDF triples with integer valued subjects, predicates, objects in Virtuoso and MonetDB's
native columnstore, and translated all the OPT pattern queries to their equivalent SQL counterpart.
We setup all the required indexes, configuration, and ``vectorized query execution'' in Virtuoso
as outlined in \cite{virtsetup,virtwp}.
For MonetDB, we created separate predicate tables with triples of only that predicate,
ordered on S-O columns as outlined in \cite{swans}, and also created an O-S index on each of these tables.
However, due to a large number of predicates in the DBPedia dataset (see Table \ref{tbl:datasets}),
MonetDB failed to create a separate table for each predicate. Hence we loaded the entire DBPedia
data in one 3-column table, and created four indexes PSO, POS, SPO, OPS on that table -- same as what
LBR and Virtuoso use. Thus we ensured that both Virtuoso and MonetDB have an optimized setup.
Current published sources of RDF-3X \cite{rdf3x} and TripleBit \cite{triplebit} cannot process OPT queries.
GH-RDF3X\footnote{\scriptsize{\url{https://github.com/gh-rdf3x/gh-rdf3x}}}, a third party system built using
sources of RDF-3X for OPT pattern support, could not correctly process many OPT queries.

\textbf{Datasets and Queries:}
We used three popular RDF datasets,
(1) Lehigh University Benchmark (LUBM) --
a synthetically generated dataset over 10,000 universities using their data generator program,
(2) UniProt \cite{uniprot} -- a real life protein network, and
(3) DBPedia (English) 2014 \cite{dbpedia} -- a real life RDF dataset created from Wikipedia.
The characteristics of these datasets are given in Table \ref{tbl:datasets}.

\begin{table}[h]
\small{
  \begin{center}
    \begin{tabular}{|p{1.1cm}|p{1.6cm}|p{1.4cm}|p{0.8cm}|p{1.4cm}|}
    \hline
    Datasets & \#triples & \# S & \# P & \# O \\
    \hline
     LUBM & 1,335,081,176 & 217,206,845 & 18 & 161,413,042 \\
     UniProt & 845,074,885 & 147,524,984 & 95 & 128,321,926 \\
     DBPedia & 565,523,796 & 29,747,387 & 57,453 & 153,561,757 \\
    \hline
    \end{tabular}
  \end{center}
}
\caption{Dataset characteristics} \label{tbl:datasets}
\end{table}
For a competitive evaluation of the three systems,
we used a mix of \textit{well-designed} OPT queries with varying degrees of selectivity,
complexity, and running times.
The LUBM data generator is popular for scalability analysis
\cite{bitmatwww10,triad,hexastore,triplebit}.
However, its queries do not cover OPT patterns well.
Therefore, to get a good coverage,
we used UniProt queries from \cite{uniq2,uniq}, and LUBM queries
from \cite{lubmq, bitmatwww10}, and introduced OPTIONAL patterns in them
by studying the Ontology and graph structure. 
Similarly, we used DBPedia queries from \cite{dbpediaspbench}
by removing union and filter clauses wherever necessary,
because the main focus of our paper is on the nested BGP-OPT queries.
Note that our approach is similar to that of other research systems,
which focus on specific SPARQL components
\cite{bitmatwww10,triad,rdf3x,triplebit,gstore}.
The current RDF benchmarks suffer from limitations as
highlighted in \cite{ibmsigmod11}. Consequently, several systems
use generated queries for specific SPARQL constructs.
We evaluated all three systems over the same OPT queries,
and they are given in Appendix \ref{apdx:queries}.

\begin{table*}[t!]
\centering
{\small
\begin{tabular}{p{0.8cm} p{0.8cm} p{0.8cm} p{1.1cm} p{1cm} p{1cm} p{1.5cm} p{1.5cm} p{1.5cm} p{1.5cm} p{1.6cm}}
\hline
 & $T_{init}$ (\textit{LBR}) & $T_{prune}$ (\textit{LBR}) & $T_{total}$ (\textit{LBR}) & $T_{Virt}$ & $T_{Monet}$ & \#initial triples & \#triples aft pruning &
  \#total results & \#results with nulls & best-match reqd? \\
\hline
Q1 & 5.88 & 3.95 & \textbf{32.69} & 104.67 & >30min & 649,375,261 & 61,662,975 & 10,448,905 & 336,455 & No \\
Q2 & 22.63 & 7.33 & \textbf{122.75} & 197.84 & >30min & 758,743,140 & 157,571,451  & 226,641 & 8449 & No \\
Q3 & 7.55 & 6.21 & \textbf{140.02} & 436.91 & >30min & 631,261,274 & 77,041,410 & 32,828,280 & 0  & No \\
Q4 & 0.38 & 0.86 & 1.29 & \textbf{0.015} & 1466.14 & 352,340,574 & 1,701,020 & 11 & 6 & Yes \\
Q5 & 0.36 & 0.85 & 1.26 & \textbf{0.018} & 1506.9 & 352,340,566 & 1,700,979 & 10 & 3 & Yes \\
Q6 & 0.57 & 0.54 & 1.22 & \textbf{0.03} & 42.71 & 438,912,504 & 1,700,913 & 7 & 0 & No \\
\hline
\end{tabular}
}
\caption{Query proc. times (in seconds, warm cache, best times boldfaced) -- LUBM 1.33 billion triples}\label{tbl:eval1}
\end{table*}

\begin{table*}[t!]
\centering
{\small
\begin{tabular}{p{0.8cm} p{0.8cm} p{0.8cm} p{1.1cm} p{1cm} p{1cm} p{1.5cm} p{1.5cm} p{1.5cm} p{1.5cm} p{1.6cm}}
\hline
 & $T_{init}$ (\textit{LBR}) & $T_{prune}$ (\textit{LBR}) & $T_{total}$ (\textit{LBR}) & $T_{Virt}$ & $T_{Monet}$ & \#initial triples & \#triples aft pruning
 & \#total results & \#results with nulls & best-match reqd? \\
\hline
Q1 & 2.53 & 1.4 & \textbf{8.2} & 13.1 & 10.8 & 546,188,591 & 13,889,249 & 917,773 & 130,161 & No \\
Q2 & 1.6 & 0 & \textbf{1.6} & 16.68 & 1.79 & 162,444,409 & 0  & 0 & 0 & No \\
Q3 & 0.89 & 0.73 & \textbf{3.25} & 7.94 & 6.57 & 38,089,712 & 11,196,191 & 1,009,371 & 1,001,134  & No \\
Q4 & 0.83 & 0.04 & \textbf{3.61} & 39.7 & 23.71 & 16,757,285 & 6,079,367 & 6,079,367 & 6,079,367 & No \\
Q5 & 0.78 & 0.47 & 1.44 & \textbf{0.49} & 1.88 & 49,696,660 & 16,258,093 & 5625 & 5625 & No \\
Q6 & 1.22 & 0.73 & 2.8 & \textbf{1.55} & 1.81 & 42,463,971 & 14,503,826 & 98,842 & 73,672 & No \\
Q7 & 2.38 & 0.48 & 4.04 & 3.37 & \textbf{2.92} & 34,666,636 & 18,833,950 & 272,822 & 1 & No \\
\hline
\end{tabular}
}
\caption{Query proc. times (in seconds, warm cache, best times boldfaced) -- UniProt 845 million triples}\label{tbl:eval2}
\end{table*}

\begin{table*}[t!]
\centering
{\small
\begin{tabular}{p{0.8cm} p{0.8cm} p{0.8cm} p{1.1cm} p{1cm} p{1cm} p{1.5cm} p{1.5cm} p{1.5cm} p{1.5cm} p{1.6cm}}
\hline
 & $T_{init}$ (\textit{LBR}) & $T_{prune}$ (\textit{LBR}) & $T_{total}$ (\textit{LBR}) & $T_{Virt}$ & $T_{Monet}$ & \#initial triples & \#triples aft pruning 
 & \#total results & \#results with nulls & best-match reqd? \\
\hline
Q1 & 1.19 & 0.19 & \textbf{3.36} & 6.65 & >10min & 22,460,545 & 3,165,560 & 515,003 & 446,204 & No \\
Q2 & 0.025 & 0 & 0.025 & \textbf{0.011} & >10min & 2,231,035 & 0 & 0 & 0 & No \\
Q3 & 0.036 & 0 & 0.036 & \textbf{0.017} & >10min & 15,799,901 & 0 & 0 & 0  & No \\
Q4 & 0.63 & 0.15 & 0.81 & \textbf{0.15} & >10min & 13,493,622 & 501,240 & 4919 & 4917 & No \\
Q5 & 0.37 & 0.04 & 0.44 & \textbf{0.023} & 0.68 & 8,624,878 & 94,336 & 5330 & 22 & No \\
Q6 & 0.4 & 0.06 & 0.48 & \textbf{0.02} & 0.83 & 50,439,668 & 48 & 36 & 36 & No \\
\hline
\end{tabular}
}
\caption{Query proc. times (in seconds, warm cache, best times boldfaced) -- DBPedia 565 million triples}\label{tbl:eval3}
\end{table*}

\textbf{Evaluation Metrics:}
We used the following metrics for evaluation:
(1) Time required for the \texttt{init} process in Alg \ref{alg:qproc} ($T_{init}$), i.e.,
the time to load the BitMats associated with all the triple patterns in a query.
(2) Time required for \texttt{prune\_triples} (Alg \ref{alg:prune}) ($T_{prune}$).
(3) Total query execution time (warm cache)
for each system averaged over 5 runs, i.e., end-to-end \textit{clock-time}
($T_{total}$, $T_{Virt}$, $T_{Monet}$ for LBR, Virtuoso, and MonetDB respectively).
We ran each query 6 times by discarding the first runtime to warm up the caches.
Also, for reporting this time fairly,
we redirected output results of all the three systems to \texttt{/dev/null},
to eliminate any overhead of I/O latency in writing the results to the disk.
Thus this reported time is the core query processing time spent by each system.
$T_{total}$ of LBR is $T_{init} + T_{prune} + T_{multiway}$.
$T_{multiway}$ (Alg \ref{alg:finalres}) is not explicitly shown in the evaluation
for conciseness, as it can be computed by simply
subtracting $T_{init} + T_{prune}$ from $T_{total}$.
(4) Initial number of triples -- the sum of triples matching each triple pattern in the query
before the \texttt{init} and \texttt{prune\_triples} procedures.
(5) Sum of triples left in all the BitMats after \texttt{prune\_triples}.
(6) Number of final results.
(7) Number of results, which have one or more NULL valued bindings, i.e.,
their result rows do not have all the variables bound due to the left-outer-joins in the query.
These latter four metrics help us highlight the selectivity properties of the queries.
(8) Whether \texttt{nullification} and \texttt{best-match} operations were required for LBR
(see Section \ref{sec:qproc}).
Query processing times over all the three datasets are given in Tables \ref{tbl:eval1}, \ref{tbl:eval2},
and \ref{tbl:eval3}.

Note that the LBR system does not run in ``server mode'' as Virtuoso and MonetDB do,
and it does not use any sophisticated cache management techniques of its own.
It may only benefit from the underlying operating system kernel's file-system cache management.

\subsection{Analysis of the Evaluation} \label{sec:analres}
From Tables \ref{tbl:eval1}, \ref{tbl:eval2}, and \ref{tbl:eval3},
we can see that for queries with low selectivity,
LBR's technique clearly excels over Virtuoso and MonetDB.

Queries Q1--Q3 of LUBM need to access more than 50\% of the data (see ``\#initial triples'' column
for the respective queries). These queries have a large number of triple patterns, have multiple OPT patterns,
and generate a large number of results. Notably, these three queries
have cyclic graphs of join-variables (GoJ), but only one join variable in each slave supernode.
So LBR does not need to use nullification and best-match operations
(see Section \ref{sec:semijloj} and Lemma \ref{lemma:cyclenobestm}).
For these three queries LBR gives several fold better performance than Virtuoso and MonetDB.

Queries Q4--Q6 of LUBM are short running, simpler queries with one OPT pattern.
A closer look at these queries shows us that
they have one or two highly selective triple patterns, with fixed values in
predicate and object positions, and joins on S-S positions.
This helps Virtuoso by way of fast merge-joins. Notice the column ``\#triples aft pruning'', which shows
that these queries actually deal with a very small fraction of the entire data -- about a million triples
out of more than a billion triple dataset, and they produce very small number of results.
Notably, for these queries, while Virtuoso performs much better, MonetDB suffers a lot. We conjecture that
for such queries with highly selective \textit{master} triple patterns, Virtuoso probably
uses the idea of basic reordering of the inner and left-outer joins. 
Virtuoso's ``explain'' tool revealed that it uses a combination of hash and bloom
filters to join the selective master with a non-selective slave.
Q4 and Q5 have a cycle in their respective GoJ, and since they also have more than one jvars in
their slave supernodes, LBR uses nullification and best-match operations to remove any
subsumed results (ref Section \ref{sec:cyclic}).

Among UniProt queries, for Q1--Q4, LBR performs much better.
These queries have multiple nestings of BGP-OPT patterns, and do not have highly selective triple
patterns in them.
Notably, for Q4 a \texttt{semi-join} between master and slave removes all the bindings in the slave.
Virtuoso however could not exploit this situation (as it did for LUBM Q4--Q6), because this master TP
is not very selective and has only predicate position fixed. LBR's pruning method with semi-joins shows
a benefit here. For Q2, LBR's \texttt{init} procedure with active pruning
detects empty results of the query much earlier, and abandons further query processing
(recall our ``simple optimization'' from Section \ref{sec:qproc}). Whereas,
Virtuoso and MonetDB detect empty results much later in their query processing.
All seven UniProt queries are \textit{acyclic}.

Similar to our previous observations, in case of DBPedia queries, 
for a query like Q1, which needs to handle larger amount of data, and produces more results,
LBR comes out as a clear winner.
Q2--Q6 access a much smaller fraction of the data, and generate a small set of results.
In case of Q2--Q3, LBR's initialization with active pruning detects an empty set of results
early on and abandons further query processing.
MonetDB suffers badly for Q1--Q4. We conjecture that this is because it was unable to create
57,453 distinct predicate tables. We successfully created four types of indexes on MonetDB's DBPedia
table, but its documentation revealed to us that MonetDB does not always honor
user created indexes; it relies on its own policies of indexing and managing the data.
All six DBPedia queries are acyclic.

In general, observe that for low selectivity queries, Q1--Q3 (LUBM), Q1--Q4 (UniProt), and Q1 (DBPedia),
LBR prunes a significant portion of the initial triples
(see columns ``\#initial triples'' and ``\#triples aft pruning''), in a relatively small
amount of time ($T_{prune}$ is a very small portion of $T_{total}$). We believe that our novel technique 
of jvar-tree traversal (see Section \ref{sec:semijloj}) along with the compressed index structure
and \textit{fold-unfold} operations (see Section \ref{sec:qproc})
make the pruning procedure efficient.
Also please notice that LBR processes long running queries \textit{several} seconds or minutes faster than Virtuoso
and MonetDB $(T_{Virt} - T_{total})$. Virtuoso does better for short running queries, 
but the difference $(T_{total} - T_{Virt})$ is just about \textit{a second}.
We feel that for such queries, in the future, LBR can probably use a more sophisticated cache management
technique by running in server mode.
We trust that for long running queries, the essence of LBR's optimization technique
gets highlighted.

To summarize, our evaluation
shows that LBR works much better than the contemporary columnstores
for queries with lower selectivity and
higher complexity, e.g., queries Q1--Q3 (LUBM), Q1-Q4 (UniProt), and Q1 (DBPedia).
The geometric mean of the presented queries for each dataset was as follows --
for UniProt, 3.05 sec (LBR), 5.61 sec (Virtuoso), 4.35 (MonetDB),
for LUBM, 10.18 sec (LBR) and 2.04 sec (Virtuoso), and for DBPedia 0.28 sec (LBR) and 0.07 sec (Virtuoso).
The geometric mean of Virtuoso is lower than LBR for LUBM and DBPedia due to short running queries
like Q4--Q6 (LUBM) and Q2--Q6 (DBPedia).
We did not compute the geometric mean of MonetDB on LUBM and DBPedia as it took very long time
for some of the queries. 

\textbf{Index Sizes:}
The on disk size of LBR indexes, i.e., $2|V_p| + |V_s| + |V_o|$ BitMats,
with our enhancements of ``hybrid compression'' technique 
is 20 GB, 32 GB, and 41 GB for DBPedia, UniProt, and LUBM datasets respectively.
But note that we do not need to load all these indexes in memory.
For any given query, we just load the BitMats associated
with the triple patterns in the query, which are typically a
smaller fraction of the total indexes. We ran our experiments
on a machine with 8 GB memory, which could fit the
BitMats associated with the triple patterns in the queries.
For Virtuoso, the on-disk size of the stored data and indexes 
was 5.5 GB, 9.5 GB, and 11 GB for DBPedia, UniProt, and LUBM datasets respectively,
and for MonetDB it was 32 GB, 23 GB, 35 GB for DBPedia, UniProt, and LUBM respectively.

\section{Related Work} \label{sec:relwork}
While the SPARQL OPT-free basic graph pattern (BGP) queries and their optimization have enjoyed quite bit of
attention from the research community in the past few years, the discussion
about OPTIONAL pattern queries has mostly remained theoretical.
Previous work, such as \cite{arenas,iswc14, perez2, schmidt},
has extensively analyzed the semantics of well-designed OPT patterns,
as they have high occurrence \cite{usewod11,swim}, 
bear desirable properties for the complexity analysis of the evaluation,
and remain unaffected by the disparity between SPARQL and SQL
algebra about joins over NULLs.
Our query graph of supernodes (GoSN) is reminiscent of well-designed pattern trees (WDPT)
\cite{perez,letelier2}, but WDPTs are undirected and unordered, whereas GoSN is directed, and
establishes an order among the patterns (\textit{master-slave}, \textit{peers}),
which is an integral part of our optimization technique.
While the work on WDPTs focuses on the analysis of containment
and equivalence of well-designed patterns and identifying the tractable
components of their evaluation, we use GoSN to focus on the practical
aspects of OPT pattern evaluation. 

From the practical aspects of the optimization of SQL left-outer-joins,
most prominently, Galindo-Legaria, Rosenthal \cite{galindosigmod,galindo-legaria1,galindo-legaria2}
and Rao et al \cite{rao2, rao1} have proposed ways of achieving it through reordering
inner and left-outer joins. Rao et al have proposed \textit{nullification} and \textit{best-match}
operators to handle inconsistent variable bindings and subsumed results respectively (see Section \ref{sec:bestm}).
In their technique, nullification and best-match are required
for each reordered query, as the minimality of tuples is not guaranteed (see Lemma \ref{lemma:min}).
They do not use methods like semi-joins to prune the candidate tuples.
Bernstein et al and Ullman \cite{semij2,semij1,ullman} have proved the properties of \textit{minimality}
for \textit{acyclic inner-joins} only.
Through LBR, we have taken a step forward by extending these semi-join properties in the context
of SPARQL OPT patterns, by analyzing the graph of jvar-nodes (GoJ), and finding ways to avoid
overheads like \textit{nullification} and \textit{best-match} operations.

For inner-join optimization, RDF engines like
TriAD \cite{triad}, RDF-3X \cite{rdf3x}, gStore \cite{gstore}
take the approaches like graph summarization, sideways-information-passing etc
for an early pruning of triples.
Systems like TripleBit \cite{triplebit} use a variable length bitwise encoding of RDF triples, and a query plan generation
that favors queries with ``star'' joins, i.e., many triple patterns joining over a single variable.
RDF engines built on top of commercial databases such as DB2RDF \cite{ibmsigmod13} propose
creation of \textit{entity-oriented} flexible schemas and better data-flow techniques
through the query plan to improve the performance of ``star'' join queries.
Along with this, there are distributed RDF processing engines such as H-RDF-3X \cite{hrdf3x} and SHARD \cite{shard}.
While many of these engines mainly focus on efficient indexing of RDF graphs,
BGP queries, and exploiting ``star'' patterns in the queries,
we have focused on the OPT patterns that may have multiple S-O joins, which cannot exploit
the benefits of ``star'' query optimizers.

Our indexes and pruning procedure resemble the BitMat system \cite{bitmatwww10}.
However, BitMat only handles BGP queries. Their query graph cannot capture
the structural aspects of a nested BGP-OPT query, and their query processing 
does not honor any left-outer-joins. Additionally, while using their index structure
as a base in our system, we have enhanced it further to reduce  as much as
40\% of the index size compared to their original method \cite{bitmatsrc}.

\section{Conclusion}
In this paper, we proposed \textit{Left Bit Right} (LBR), for optimizing
SPARQL OPTIONAL pattern queries.
We proposed a novel concept of a query graph of \textit{supernodes} to capture the structure
of a nested OPT query.
We proposed optimization strategies -- first of a kind, to the best of our knowledge -- that
extend the previously known properties of acyclicity, minimality, nullification, and best-match.
We presented LBR's evaluation in comparison with the two mainstream RDF columnstores,
Virtuoso and MonetDB, and showed that LBR's technique works much better
for low-selectivity OPT queries with multiple OPT patterns, and
for highly selective simpler queries, it gives at par performance with the other systems.
In the future, we plan to extend our query processor to handle other SPARQL constructs
such as unions, filters, and we intend to investigate methods
for better cache management especially for short running queries.

\section*{Acknowledgements}
We are grateful to Dr. Sujatha Das Gollapalli and Dr. Sudeepa Roy
for their valuable inputs, and the three anonymous reviewers for their detailed feedback,
which significantly helped us in improving the presentation of our paper.


\appendix
\section{Proofs}
\subsection{Lemma \ref{lemma:min}} \label{apdx:lemma:min}
\begin{proof}
Let nullification be required with a minimal set of triples in each TP in a query.
This means that there are one or more variable bindings and triples that will be
removed as a result of an inner or left-outer-join
that caused NULL values for the respective variable bindings.
This is a contradiction to our assumption of minimality.

Let us assume that there are subsumed results with a minimal set of triples,
and best-match is required to remove them.
Let result $r_2$ be subsumed by $r_1$ ($r_2 \sqsubset r_1$), and
$non\mbox{-}null(r_1)$ and $non\mbox{-}null(r_2)$ be the non-NULL variable bindings
in $r_1$ and $r_2$ respectively.
Let $S_1 = non\mbox{-}null(r_1) \setminus non\mbox{-}null(r_2)$
and $S_2 = non\mbox{-}null(r_1) \cap non\mbox{-}null(r_2)$.
Per the definition of subsumption,
$non\mbox{-}null(r_2) \subset non\mbox{-}null(r_1)$. 
This means that $S_1$ are the variable bindings
contributed by a ``slave'' TP, which is why they could be set to NULL in $r_2$
(see the definition of master-slave in Section \ref{sec:nomen}).
This also means that $S_2$ variable bindings had corresponding
bindings of variables in $S_1$ in $r_1$ but not in $r_2$, implying that
those $S_1$ bindings and the respective triples were \textit{removed}
in the process of joins. This is a contradiction to our assumption of minimality of triples.
\end{proof}

\subsection{Lemma \ref{lemma:acyclic}} \label{apdx:lemma:acyclic}
\begin{proof}
Consider a query with three join variables, ?a, ?b, ?c, that form a three-cycle
in GoJ. Let us assume that the respective GoT of the query
is acyclic. An edge between ?a and ?b indicates that there exists a triple pattern $tp_1$, which has both
?a and ?b in it. Similarly there must be a $tp_2$ with join variables ?b and ?c, and $tp_3$
with join variables ?a and ?c.
Now, $tp_1$ and $tp_2$ join on ?b, $tp_1$ and $tp_3$ join on ?a, and $tp_2$ and $tp_3$
join on ?c. This indicates a cycle among the TPs which is a contradiction. The same analysis
can be extended to a join-variable cycle of more than three length.

Now consider a GoJ with no cycles, i.e., it is a tree. Consider that there
exists a cycle involving TPs $tp_1...tp_n$ in GoT, such that each $tp_i$ and $tp_{i+1}$,
$1 \leq i < n$,
join over join variable $?j_i$, and $tp_1$ and $tp_n$ join over $?j_n$. This means that
jvar-node graph has an edge between every pair of jvar-nodes $?j_i$ and $j_{i+1}$, $1 \leq i < n$,
and there is an edge between $?j_n$ and $?j_1$. This indicates a cycle in the jvar-node graph,
which is a contradiction.
\end{proof}

\subsection{Lemma \ref{lemma:optmin}} \label{apdx:lemma:optmin}
\begin{proof}
Let an OPT pattern be $P_k \leftouterjoin P_l$ with $P_k$ and $P_l$ as OPT-free BGPs.
Let us temporarily ignore the left-outer-join
between $P_k$ and $P_l$, and consider the two BGPs independently. $P_k$ and $P_l$ both have an
acyclic connected GoJ. 
This observation follows from the fact that GoJ of $P_k$ and $P_l$ together is acyclic, and there are 
no Cartesian products in the query. Doing a bottom-up followed by top-down pass on the induced subtrees of $P_k$
and $P_l$ \textit{independently}, with only \textit{clustered-semi-joins} for each jvar, ensures
that TPs in $P_k$ and $P_l$ are left with a minimal set of triples. This follows
from Property \ref{prop:goj} and is proved in \cite{semij2}.
Now consider $P_k \leftouterjoin P_l$. Let $\mathcal{J} = jvars(P_k) \cap jvars(P_l)$, 
and $\mathcal{J} \neq \phi$ because GoJ of $P_k \leftouterjoin P_l$ is connected.
In Algorithm \ref{alg:jvarord}, line \ref{ln:mroot} chooses one of the jvars in
$\mathcal{J}$ to be the root of $P_l$'s induced subtree. Considering
only jvars in $\mathcal{J}$ and acyclicity, this means that $P_l$ has an induced subtree
where jvar-nodes shared with its masters ($\mathcal{J}$) appear as the
\textit{ancestors} of the jvars that appear only in $P_l$, but not in $P_k$.

A bottom-up pass over jvar subtrees of $P_k$ and $P_l$
(ln \ref{ln:prunebu}--\ref{ln:prunebuend} in Alg \ref{alg:prune}),
transfers the restrictions on the variable bindings across respective TPs in $P_k$ and $P_l$.
Note that \texttt{semi-join} transfers the restrictions on variable bindings from $P_k$ to $P_l$,
and \texttt{clustered-semi-join} transfers the restrictions on bindings among peers.
Recall that we build $order_{td}$ such that induced subtree of $P_k$ is traversed before $P_l$'s
(ln \ref{ln:tdpass}--\ref{ln:slavetreeend2} in Alg \ref{alg:jvarord}).
A top-down pass on the jvar subtree of $P_k$ (ln \ref{ln:tdpasssj}--\ref{ln:tdpasssjend}
in Alg \ref{alg:prune}) leaves a \textit{minimal} set of triples in the TPs in $P_k$.
Hence a top-down pass on the jvar subtree of $P_l$ after $P_k$'s,
leaves a minimal set of triples in the TPs in $P_l$ as a \textit{ripple effect}.

This analysis can be inductively applied for any nesting of BGP and OPT patterns, as long as their GoJ
is acyclic and connected. Hence with the procedures in Algorithms \ref{alg:jvarord} and \ref{alg:prune}, we can
get a minimal set of triples in each TP in a nested acyclic BGP-OPT query.
\end{proof}

\subsection{Lemma \ref{lemma:cyclenobestm}} \label{apdx:lemma:cyclenobestm}
\begin{proof}
We need to use \textit{nullification} and \textit{best-match} if a
slave supernode has more than one jvars, because removal of bindings
of one jvar may change the bindings of another due to the \textit{ripple effect} through 
GoJ. Let us consider an OPT pattern of kind $tp_1 \leftouterjoin_{?j} (tp_2 \Join_{?j} tp_3)$.
Let $P_1 = (tp_1)$ and $P_2 = (tp_2 \Join_{?j} tp_3)$.
$P_2$ has only one jvar $?j$, that it shares with the master $P_1$.
In \texttt{prune\_triples} (Alg \ref{alg:prune}), we 
transfer the restrictions on bindings of $?j$ from $tp_1$ to $tp_2$ and $tp_3$
through a \texttt{semi-join}.
Through a \texttt{clustered-semi-join} we ensure that restrictions on bindings of $?j$ are transferred
among $tp_2$ and $tp_3$. So even if the original OPT pattern is reordered
as $(tp_1 \leftouterjoin tp_2) \leftouterjoin tp_3$,
there will not be any subsumed results.
This is because, after a \texttt{clustered-semi-join}, $tp_3$ does not have any bindings of \textit{?j}
that do not appear in $tp_2$ and vice versa.

An alternate way to look at it is:
Consider a cyclic OPT pattern where slaves have more than one jvars.
Then if a final result $r_4$ of the query is subsumed by $r_3$ ($r_4 \sqsubset r_3$),
$non\mbox{-}null(r_3) \setminus non\mbox{-}null(r_4)$ have at least one jvar
that appears \textit{only} in the slaves of the query,
i.e., $non\mbox{-}null(r_3) \setminus non\mbox{-}null(r_4)$ variables
do not appear in \textit{absolute masters} -- that is why they could be set to NULLs.
Let $S_1 = non\mbox{-}null(r_3) \setminus non\mbox{-}null(r_4)$,
$S_2 = non\mbox{-}null(r_3) \cap non\mbox{-}null(r_4)$.
Note that \textit{not all} $S_1$ can be non-jvars from slaves,
because it would mean that while variables in $S_2$ had the
respective bindings for those in $S_1$ in $r_3$, they did not find any bindings for $S_1$
in $r_4$. As per our \texttt{prune\_triples} procedure (Alg. \ref{alg:prune}),
we only prune the bindings of jvars; bindings of non-jvars
get removed as a \textit{side effect} of the pruning of jvar bindings.
So if $S_1$ has all non-jvars, this will be a contradiction.
Hence, $S_1$ would have at least one join variable that \textit{appears only in a slave},
for subsumed result $r_4$.
But for a cyclic OPT pattern (without Cartesian products)
with each slave having only one jvar, each jvar
appears in an \textit{absolute master} too. This in turn means that none of the jvars
in the query can be set to NULL. $S_1$ does not have any jvars that appear
only in slaves, which means that subsumed results cannot be generated.

Thus we do not need \textit{nullification} and \textit{best-match},
for a cyclic OPT pattern query, where each slave has only one jvar.
\end{proof}

\section{Non-well-designed patterns} \label{apdx:wdpat}
As defined by P\'{e}rez et al in \cite{perez2}, a SPARQL nested BGP-OPT pattern query $P$ is
said to be \textit{well-designed} (WD)
if for every subpattern $P' = (P_i \leftouterjoin P_j)$ of $P$, if a join variable $?j$
occurs in $P_j$ and outside $P'$, then $?j$ occurs in $P_i$ as well.
An OPT pattern that violates this condition is called a non-well-designed (NWD) pattern.

Before we discuss the evaluation of NWD patterns, it is important to take into
consideration how NULLs are treated in left-outer-joins or inner-joins. There is a
disparity between the SPARQL and SQL algebra over this, which
mainly creates a problem for NWD queries. The same NWD query evaluated over
a pure SPARQL engine, e.g., Jena, gives counter-intuitive results than a SPARQL-over-SQL engine,
e.g. Virtuoso, when there are joins over NULLs. This issue is discussed in detail with an example
in Appendix \ref{apdx:nulltreat}.
However, using the same assumption of ``null-intolerant'' (or null-rejecting) 
joins in SQL as done in the previous literature by Rao et al and Galindo-Legaria,
Rosenthal \cite{rao1, galindo-legaria2}, we give a way of simplifying NWD patterns.
In a ``null-intolerant'' join evaluation, NULL values are not matched
with anything including other NULL values.
E.g., in a left-outer-join $P_a \leftouterjoin_{?j} P_b$,
if $?j$ has NULL values in $P_b$ \textit{before the join},
they get eliminated during the left-outer-join evaluation.
NULLs are only introduced as a result of a left-outer-join for missing values in
the right hand side pattern. 

For any OPT query (WD or NWD), we first serialize
it using OPT-free BGPs, join operators $\Join$, $\leftouterjoin$, and proper parentheses;
then build a graph of supernodes (GoSN) as described in Section \ref{sec:qgrconstr1}.
If the given query is NWD, we identify all OPT-free BGPs that \textit{violate} the WD condition,
and the corresponding OPT-free BGPs with which they do a violation. E.g., if the query is
$P_x \Join (P_y \leftouterjoin P_z)$, where $P_z$ has a join variable
$?j$ that appears in $P_x$ but not in $P_y$, then we say, \textit{``$P_z$ violates WD condition
with $P_x$, and ($P_z, P_x$) is a violation pair''}. We identify all such violation
pairs in the query, and their corresponding violation supernode pairs in the GoSN. For this example
($SN_z, SN_x$) will be the violation pair of supernodes.
Temporarily ignoring the directionality of edges in GoSN, we identify
the undirected path between each such supernode violation pair. Note that
according to GoSN's construction, there is a unique undirected path between any pair of
supernodes. Next, considering the original directionality of edges
on this path, we convert any unidirectional edges into bidirectional edges. 

\begin{figure}[h]
    \centering
        \includegraphics[width=3in,height=0.7in]{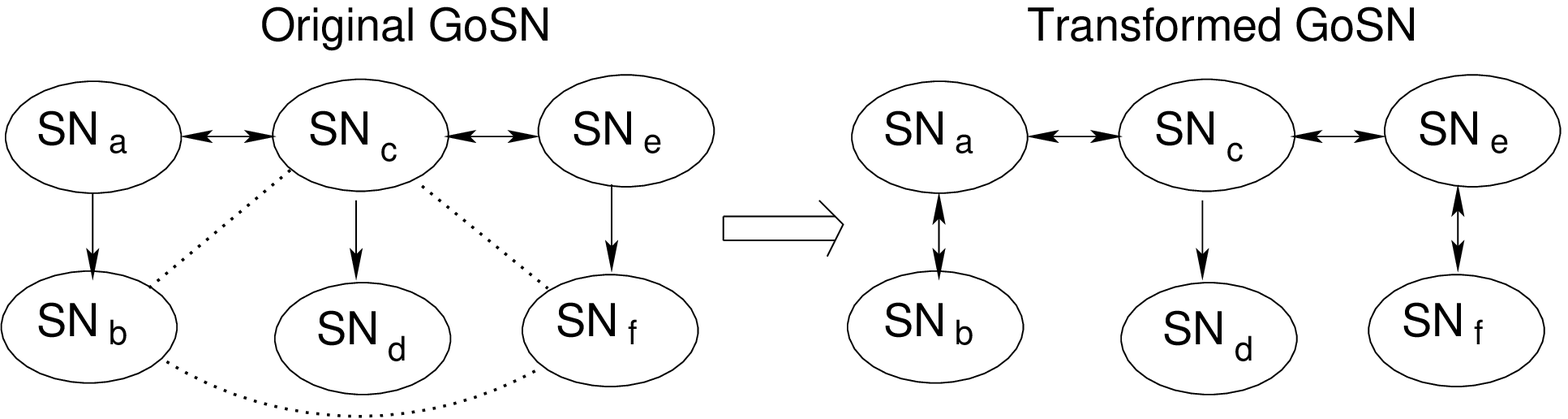}
        \caption{NWD query GoSN transformation}
    \label{fig:nwd}
\end{figure}

Conversion of unidirectional edges into bidirectional edges is continued this way until all the violation
pairs in GoSN are treated. Note that this is a monotonic process, and always converges -- we always convert
unidirectional edges into bidirectional edges, and never the other way round.
This effectively means we convert one or more left-outer-joins
in the original query into inner-joins (recall that a unidirectional edge represents a left-outer-join,
and a bidirectional edge represents inner-join).
E.g., consider a serialized query $(P_a \leftouterjoin P_b) \Join ((P_c \leftouterjoin P_d)
\Join (P_e \leftouterjoin P_f))$, where $P_b$ violates WD condition over variable $?j_1$ with
$P_c$, and $P_f$ also violates WD condition over the same variable $?j_1$ with $P_c$. In turn,
$P_b$ and $P_f$ too violate WD condition with each other over $?j_1$.
In Figure \ref{fig:nwd}, we show the original GoSN and the transformed GoSN for this query. The dotted lines
in the original GoSN indicate the pairs of respective supernodes that are in violation
of the WD condition. 

After this transformation process, we use the same query processing
techniques given in Algorithms \ref{alg:jvarord}, \ref{alg:prune}, and
\ref{alg:qproc} on this GoSN with null-intolerant join assumption.

\section{Treatment of Nulls} \label{apdx:nulltreat}
Unlike relational tables, RDF instance data does not contain
NULLs\footnote{\scriptsize{The difference between NULLs and ``blank nodes'' is elaborated
in Section \ref{sec:nomen}.}}.
Hence the issue of treatment of NULLs predominantly arises in case of
non-well-designed OPT queries.
This is because of the difference between the semantics of left-outer-joins 
between the SPARQL specifications \cite{sparql} and SQL algebra.
Borrowing the definitions from P\'{e}rez et al in \cite{perez2},
if we have a pattern $P_i \leftouterjoin P_j$ with $\Omega_{i}$ and $\Omega_{j}$ as the set of
\textit{mappings} (variable bindings) of $P_i$ and $P_j$ respectively, then a SPARQL OPT pattern is defined as:
\[
 \Omega_i \leftouterjoin \Omega_j = (\Omega_i \Join \Omega_j) \cup (\Omega_i \setminus \Omega_j)
\]
This definition allows $P_i \leftouterjoin P_j$ to have results which may have different \textit{arity}.
For example, considering the query in Figure \ref{fig:nullbest}, per above definition,
the results will be \{(:Larry), (:Julia, :Seinfeld)\}.
Note that the first result has arity 1, whereas the second one has arity 2. 

This causes a problem if we do an inner-join of these results (mappings) with
another pattern over \textit{?sitcom} (as may happen in case of
a non-well-designed query). Per SPARQL semantics of inner-join ($\Join$),
two mappings are considered \textit{join-compatible}, if they match on the variables that
are \textit{bound} in the respective
mappings. For any unbound variables, the mappings are still considered compatible.
E.g., if we join the above mappings with \{(:Friends)\} over the \textit{?sitcom} variable, the result will be
counter-intuitive -- \{(:Larry, :Friends)\}. Note that mapping (:Julia, :Seinfeld) got removed,
but (:Larry) was preserved, because it does not have an explicit NULL value for \textit{?sitcom}. 

In relational databases, evaluation of joins over NULL values has had different interpretations
in different contexts (see \cite{perez2, rao2} for a discussion).
However, for all practical purposes, most mainstream relational 
database systems assume a ``null-intolerant'' join evaluation.
Also relational algebra assumes that the two sets of mappings (tuples) that are unioned,
are \textit{union compatible}, i.e., they have the same arity and same attributes \cite{dbbook}.
Due to these differences in the join and union semantics of SQL
and SPARQL, RDF stores that are built on top of relational databases, e.g., Virtuoso, give
different results for a non-well-designed query, compared to a native SPARQL processing engine,
when joining over NULLs.

We observed that pure SPARQL processing engines, such as ARQ/Jena \cite{jena},
follow the join semantics which allow union of mappings with different arity, and a 
\textit{null-tolerant} join, whereas relational RDF stores such as Virtuoso or MonetDB follow
the SQL semantics of union-compatibility and null-intolerant joins.

\section{Conceptual Bitcube} \label{apdx:bitmat}
Borrowing the description of conceptual bitcube construction from \cite{bitmatwww10},
here we elaborate on the process of mapping the unique S, P, O values
in the RDF data to the each bitcube dimension.
As described in Section \ref{sec:bitmat}, the dimensions of the 3D bitcube of an RDF dataset are
$V_s \times V_p \times V_o$.
The unique values of S, P, O in the original RDF data are first
mapped to integer IDs, which in turn are mapped to the respective bitcube dimensions.
Let $V_{so} = V_s \cap V_o$. Set $V_{so}$ is mapped to a sequence of integers 1 to $|V_{so}|$.
Set $V_s - V_{so}$ is mapped to a sequence of integers $|V_{so}| + 1$ to $|V_s|$.
Set $V_o - V_{so}$ is mapped to a sequence of integers $|V_{so}| + 1$ to $|V_o|$,
and set $V_p$ is mapped to a sequence of integers 1 to $|V_p|$.
The common S-O identifier assignment ($V_{so}$) is for the sake of
S-O joins.

The 2D BitMats are created by slicing this bitcube along each dimension, and they 
are compressed using our ``hybrid compression'' scheme as described
in Section \ref{sec:bitmat}.
Other meta-information such as, the number of triples, and condensed representation of
all the non-empty rows and columns in each BitMat, is also stored along with them.
This information helps us in quickly determining the number of triples in each
BitMat and its selectivity without counting each triple in it,
while processing the queries.

\section{Queries} \label{apdx:queries}
{\scriptsize
\subsection{LUBM Queries}
\vspace{3mm}
\noindent PREFIX ub: $<$http://www.lehigh.edu/\textasciitilde zhp2/2004/0401/univ-bench-.owl\#$>$\\

\noindent \textbf{Q1:}
SELECT * WHERE \{\{
?st ub:teachingAssistantOf ?course .
\textbf{OPTIONAL} \{
?st ub:takesCourse ?course2 .
?pub1 ub:publication-Author ?st .
\}\}
\{
?prof ub:teacherOf ?course .
?st ub:advisor ?prof .
\textbf{OPTIONAL}
\{
?prof ub:researchInterest ?resint .
?pub2 ub:publica-tionAuthor ?prof .
\}\}\}

\noindent \textbf{Q2:}
SELECT * WHERE \{\{
?pub rdf:type :Publication . ?pub ub:pu-blicationAuthor ?st .
?pub ub:publicationAuthor ?prof . \textbf{OPTION-AL} \{
?st ub:emailAddress ?ste . ?st ub:telephone ?sttel .
\}\}\{?st ub:undergraduateDegreeFrom ?univ . ?dept ub:subOrganizationOf ?univ .
\textbf{OPTIONAL} \{ ?head ub:headOf ?dept . ?others ub:worksFor ?dept .
\}\}\{ ?st ub:memberOf ?dept . ?prof ub:worksFor ?dept .
\textbf{OPTIONAL} \{?prof ub:doctoralDegreeFrom ?univ1 . ?prof ub:research-Interest ?resint1 .
\}\}\}

\noindent \textbf{Q3:}
SELECT * WHERE \{ \{
?pub ub:publicationAuthor ?st . ?pub ub:publicationAuthor ?prof .
?st rdf:type :GraduateStudent . \textbf{OPTIONAL}\{
?st ub:undergraduateDegreeFrom ?univ1 .
?st ub:telepho-ne ?sttel .
\}\}
\{
?st ub:advisor ?prof .
\textbf{OPTIONAL} \{
?prof ub:docto-ralDegreeFrom ?univ .
?prof ub:researchInterest ?resint .
\}\}\{ ?st ub:memberOf ?dept .
?prof ub:worksFor ?dept .
?prof a ub:FullProfe-ssor .
\textbf{OPTIONAL} \{
?head ub:headOf ?dept . ?others ub:worksFor ?dept .
\}\}\}

\noindent \textbf{Q4:}
SELECT * WHERE \{
?x ub:worksFor <http://www.Departme-nt9.University9999.edu> .
?x a ub:FullProfessor . \textbf{OPTIONAL} \{
?y ub:advisor ?x .
?x ub:teacherOf ?z .
?y ub:takesCourse ?z .
\}

\noindent \textbf{Q5:}
SELECT * WHERE \{\{
?x ub:worksFor <http://www.Departme-nt0.University12.edu> .
?x a ub:FullProfessor .
\textbf{OPTIONAL} \{
?y ub:advisor ?x .
?x ub:teacherOf ?z .
?y ub:takesCourse ?z .
\}
\}

\noindent \textbf{Q6:}
SELECT * WHERE
\{
?x ub:worksFor <http://www.Depart-ment0.University12.edu> .
?x a ub:FullProfessor> .
\textbf{OPTIONAL} \{
?x ub:emailAddress ?y1 .
?x ub:telephone ?y2 .
?x ub:name ?y3 .
\}
\}

\subsection{UniProt queries}
\vspace{3mm}
\noindent PREFIX uni: $<$http://purl.uniprot.org/core/$>$\\
PREFIX schema: $<$http://www.w3.org/2000/01/rdf-schema\#$>$\\

\noindent\textbf{Q1:}
SELECT *
WHERE \{
\{
?protein rdf:type uni:Protein .
?protein uni:recommendedName ?rn .
\textbf{OPTIONAL} \{
	 ?rn uni:fullName ?name .
	?rn rdf:type ?rntype .
\}
\}
\{
	?protein uni:encodedBy ?gene .
	\textbf{OPTIONAL} \{
	?gene uni:name ?gn .
	?gene rdf:type ?gtype .
	\}
\}
\{
?protein uni:sequence ?seq .
?seq a ?stype .
\}
\}

\noindent\textbf{Q2:}
SELECT *
WHERE
\{
\{
	?a rdf:subject ?b .
	?a uni:encodedBy ?vo .
	\textbf{OPTIONAL} \{
	  ?a schema:seeAlso ?x
	\}
\}
\{
	?b a :Protein .
	?b uni:sequence ?z .
	\textbf{OPTIONAL} \{
	?b uni:replaces ?c .
	\}
	\}
\{
	?z a uni:Simple\_Sequence .
	\textbf{OPTIONAL} \{
	?z uni:version ?v .
	\}
\}
\}

\noindent\textbf{Q3:} 
SELECT *
WHERE 
\{
	\{
		?protein rdf:type uni:Protein .
		?protein uni:organism $<$http://purl.uniprot.org/taxonomy/9606$>$ .
		\textbf{OPTIONAL} \{
			?protein uni:encodedBy ?gene .
			?gene uni:name ?gname .
		\}
	\}
	\{
		?protein uni:annotation ?an .
		\textbf{OPTIONAL} \{
			?an rdf:type uni:Disease\_Annotation .
			?an schema:comment ?text .
		\}
	\}
\}

\noindent\textbf{Q4:} 
SELECT *
WHERE
\{
	?s uni:encodedBy ?seq .
	\textbf{OPTIONAL} \{
		?seq uni:context ?m .
		?m schema:label ?b .
	\}
\}

\noindent\textbf{Q5:} 
SELECT *
WHERE
\{
	\{
		?a uni:replaces ?b .
		\textbf{OPTIONAL} \{
			?a uni:encodedBy ?gene .
			?gene uni:name ?name .
			?gene rdf:type uni:Gene .
		\}
	\}
	\{
		?b rdf:type uni:Protein .
		?b uni:modified ``2008-01-15'' .
		\textbf{OPTIONAL} \{
			?b uni:sequence ?seq .
			?seq uni:memberOf ?m .
		\}
	\}
\}

\noindent\textbf{Q6:} 
SELECT *
WHERE
\{
	\{
		?protein a uni:Protein .
		?protein uni:organism $<$http://purl.uniprot.org/taxonomy/9606$>$ .
		\textbf{OPTIO-NAL} \{
			?protein uni:annotation ?an .
			?an a uni:Natural\_Variant\_Anno-tation .
			?an schema:comment ?text .
		\}
	\}
	\{
		?protein uni:sequence ?seq .
		?seq rdf:value ?val .
	\}
\}

\noindent\textbf{Q7:} 
SELECT *
WHERE
\{
	?protein a uni:Protein .
	?protein uni:anno-tation ?an .
	?an a uni:Transmembrane\_Annotation .
	\textbf{OPTIONAL} \{
		?an uni:range ?range .
		?range uni:begin ?begin .
		?range uni:end ?end .
	\}
\}

\subsection{DBPedia queries}
\vspace{3mm}
\noindent
PREFIX dbpedia: $<$http://dbpedia.org/resource/$>$\\
PREFIX dbpowl: $<$http://dbpedia.org/ontology/$>$\\
PREFIX dbpprop: $<$http://dbpedia.org/property/$>$\\
PREFIX dbpyago: $<$http://dbpedia.org/class/yago/$>$\\
PREFIX dbpcat: $<$http://dbpedia.org/resource/Category/$>$\\
PREFIX rdf: $<$http://www.w3.org/1999/02/22-rdf-syntax-ns\#$>$\\
PREFIX rdfs: $<$http://www.w3.org/2000/01/rdf-schema\#$>$\\
PREFIX foaf: $<$http://xmlns.com/foaf/0.1/$>$\\
PREFIX geo: $<$http://www.w3.org/2003/01/geo/wgs84\_pos\#$>$\\
PREFIX owl: $<$http://www.w3.org/2002/07/owl\#$>$\\
PREFIX xsd: $<$http://www.w3.org/2001/XMLSchema\#$>$\\
PREFIX skos: $<$http://www.w3.org/2004/02/skos/core\#$>$\\
PREFIX georss: $<$http://www.georss.org/georss/$>$\\

\noindent\textbf{Q1:} 
SELECT *
WHERE
\{
	\{
		?v6 a dbpowl:PopulatedPlace .
		?v6  dbpowl:abstract ?v1 .
		?v6 rdfs:label ?v2 .
		?v6 geo:lat ?v3 .
		?v6  geo:long ?v4 .
		\textbf{OPTIONAL} \{
			?v6 foaf:depiction ?v8 .
		\}
	\}
	\textbf{OPTIONAL} \{
		?v6 foaf:homepage ?v10 .
	\}
	\textbf{OPTIONAL} \{
		?v6 dbpowl:-populationTotal ?v12 .
	\}
	\textbf{OPTIONAL} \{
		?v6 dbpowl:thumbnail ?v14 .
	\}
\}

\noindent\textbf{Q2:} 
SELECT *
WHERE 
\{
	?v3 foaf:page ?v0 .
	?v3 a dbpowl:Soccer-Player .
	?v3 dbpprop:position ?v6 . 
	?v3 dbpprop:clubs ?v8 .
	?v8 dbpowl:capacity ?v1 .
	?v3 dbpowl:birthPlace ?v5 .
	\textbf{OPTIONAL} \{
		?v3 dbpowl:number ?v9 .
	\}
\}

\noindent\textbf{Q3:} 
SELECT *
WHERE
\{
	?v5 dbpowl:thumbnail ?v4 .
	?v5 rdf:type dbpowl:Person .
	?v5 rdfs:label ?v .
	?v5 foaf:page ?v8 .
	\textbf{OPTIONAL} \{
		?v5 foaf:homepage ?v10 .
	\}
\}

\noindent\textbf{Q4:} 
SELECT *
WHERE 
\{
	\{
		?v2 a dbpowl:Settlement .
		?v2	rdfs:label ?v .
		?v6 a dbpowl:Airport .
		?v6 dbpowl:city ?v2 .
		?v6 dbpprop:iata ?v5 .
		\textbf{OPTIONAL} \{
			?v6 foaf:homepage ?v7 .
		\}
	\}
	\textbf{OPTIONAL} \{
		?v6 dbpprop:nativename ?v8 .
	\}
\}

\noindent\textbf{Q5:} 
SELECT *
WHERE
\{
	?v4 skos:subject ?v .
	?v4 foaf:name ?v6 .
	\textbf{OPTIONAL} \{
		?v4 rdfs:comment ?v8 .
	\}
\}

\noindent\textbf{Q6:} 
SELECT *
WHERE \{ 
	?v0 rdfs:comment ?v1 .
	?v0 foaf:page ?v .
	\textbf{OPTIONAL} \{
		?v0 skos:subject ?v6 .
	\}
	\textbf{OPTIONAL} \{
		?v0 dbpprop:industry ?v5 .
	\}
	\textbf{OPTIONAL} \{
		?v0 dbpprop:location ?v2 .
	\}
	\textbf{OPTIONAL} \{
		?v0 dbpprop:locationCountry ?v3 .
	\}
	\textbf{OPTIONAL} \{
		?v0 dbpprop:locationCity ?v9 .
		?a dbpprop:manufacturer ?v0 .
	\}
	\textbf{OPTIONAL} \{
		?v0 dbpprop:products ?v11 .
		?b  dbpprop:model ?v0 .
	\}
	\textbf{OPTIONAL} \{
		?v0 georss:point ?v10 .
	\}
	\textbf{OPTIONAL} \{
		?v0 rdf:type ?v7 .
	\}
\}
}
\end{document}